\documentclass[a4paper,11pt]{article}
\usepackage{tabularx} 
\usepackage{float}
\usepackage{xcolor}
\usepackage{amssymb}
\usepackage{amsmath}  
\usepackage{graphicx} 
\usepackage[margin=1in,letterpaper]{geometry} 
\usepackage{subcaption}
\pdfoutput=1

\usepackage{jcappub}
\usepackage{lineno}
\usepackage{makecell}

\usepackage[T1]{fontenc} 

\begin{document}

\date{\today}

\title{Radio Lines from accreting Axion Stars}

\author[a]{Dennis Maseizik,}
\author[b]{Sagnik Mondal,}
\author[c]{Hyeonseok Seong}
\author[a]{and G{\"u}nter Sigl}


\affiliation[a]{II. Institut für Theoretische Physik, Universität Hamburg, Luruper Chaussee 149, 22761 Hamburg, Germany}
\affiliation[b]{Department of Physics, University of Maryland, College Park, MD 20742, USA}
\affiliation[c]{Deutsches Elektronen-Synchrotron DESY, Notkestr. 85, 22607 Hamburg, Germany}

\emailAdd{dennis.maseizik@desy.de}
\emailAdd{sagnik@umd.edu}
\emailAdd{hyeonseok.seong@desy.de}
\emailAdd{guenter.sigl@desy.de}

\abstract{
Axion-like particles, which we call axions, can compose the missing dark matter and may form substructures such as miniclusters and axion stars.
We obtain the mass distributions of axion stars derived from their host miniclusters in our galaxy and find a significant number of axion stars reaching the \textit{decay mass}, the critical mass set by the axion-photon coupling.
Axion stars that have reached the decay mass can accrete surrounding axions either via or directly from their host miniclusters, subsequently converting them into radio photons through parametric resonance.
We demonstrate that this accretion provides observable signals by proposing two scenarios: 1)~external accretion of background dark matter occurring via miniclusters, and 2)~internal accretion of isolated systems occurring directly from the minicluster onto its core.
The emitted radio photons are nearly monochromatic with energies around the half of the axion mass.
The radio-line signal emanating from such axion stars provides a distinctive opportunity searching for axions, overcoming the widespread radio backgrounds.
We estimate the expected radio-line flux density to constrain the axion-photon coupling $g_{a\gamma\gamma}$ at each axion mass and find that the resultant line flux density is strong enough to be observed in radio telescopes such as LOFAR, FAST, ALMA, and upcoming SKA.
We can constrain the axion-photon coupling down to $g_{a\gamma\gamma} \simeq 10^{-12} - 10^{-11}\,{\rm GeV}^{-1}$, reaching even $10^{-13}\,{\rm GeV}^{-1}$ depending on the accretion models of axion stars, over an axion mass range of $m_a\simeq 10^{-7} - 10^{-2}\,{\rm eV}$.
From a different perspective, this radio-line signal could be a strong hint of an axion at the corresponding mass and also of axion stars within our galaxy.
}

\makeatletter
\def\@fpheader{}
\makeatother

\begin{center}
    \hfill  DESY-24-140
\end{center}

\maketitle

\section{Introduction}
Axion-like particles, commonly just termed axions, are light scalar particles that have been conjectured to compose the missing dark matter in our universe.
Axion dark matter may form substructures such as the galactic NFW dark matter halo, miniclusters (MCs)~\cite{Hogan:1988mp,vaquero_early_2019}, and solitonic cores inside the MCs due to gravitational interaction \cite{Schive:2014dra,Schive:2014hza,veltmaat_formation_2018}.
The solitonic cores in the context of axion dark matter are often referred to as axion stars (ASs)~\cite{Kolb:1993zz,Visinelli:2017ooc,eggemeier_formation_2019,chen_new_2021,Gorghetto:2024vnp,Chang:2024fol}.
We focus on the stable AS configurations for which the gravitation, quantum pressure and self-interactions are balanced and make the ASs stable against perturbations.
Since the ASs on this stable branch of soliton solutions have lower densities compared to the solutions on the unstable branch, these ASs are also called dilute ASs.
For stable AS configurations, the gravitational force is dominant over the weak attractive self-interaction of axions, which we consider in this work.

The ASs can decay into photons if the axion couples to the photons through
\begin{align}
    \mathcal{L}\supset \frac{g_{a\gamma\gamma}}{4}aF_{\mu\nu}\tilde{F}^{\mu\nu},
    \label{eq:axion-photon-Lagrangian}
\end{align}
where $g_{a\gamma\gamma}$ is a constant representing the coupling strength, $a$ the axion field, $F_{\mu\nu}$ the electromagnetic field strength tensor, and $\tilde{F}^{\mu\nu} = \epsilon^{\mu\nu\rho\sigma}F_{\rho\sigma}/(2\sqrt{-g})$ its dual.
This coupling induces important phenomenological consequences for the ASs, which have drawn considerable attention and have been studied rather recently \cite{Iwazaki:2014wta,Raby:2016deh,Hertzberg:2018zte,levkov_radio-emission_2020,hertzberg_merger_2020,Amin:2021tnq,Bai:2021nrs,witte_transient_2023,Eby:2024mhd} (see \cite{tkachev_fast_2015,Caputo:2018vmy,edwards_transient_2021,Witte:2021arp,Carenza_2021,Millar:2021gzs,Buen-Abad:2021qvj,Sun:2021oqp,Dev:2023ijb,Sun:2023gic,Todarello:2023xuf} for further aspects of axion-photon conversion, and even \cite{Chen:2024aqf,Arza:2024iuv} for the dark-photon counterparts).
Especially, if one photon can stimulate the decay of an axion before it exits the AS, as summarized in Eq.~\eqref{Hertzbergconst1}, the cascading decay of axions develops into a resonant emission of photons, generally referred to as parametric resonance.
Since the energy of resultant photons is strongly centered around half of the axion mass, i.e., $f = m_a/(4\pi) \simeq 120\,{\rm MHz}\,(m_a/\mu{\rm eV})$, this line signal provides a possibility to be observed against the radio background.
Here the relevant axion mass is of $\mathcal{O}({\rm neV} - {\rm meV})$.

In this work, we consider the accretion of ASs and ensuing photon signals from ASs that satisfy the parametric resonance condition in Eq.~\eqref{Hertzbergconst1}.\footnote{Preliminary results were presented in \cite{Sigl:2024pxq}.}
This condition on the axion-photon coupling can be rephrased in terms of the AS mass, which leads to a critical mass as given in Eq.~\eqref{eq:M_Decay_Gaussian}.
We refer to this critical mass as \textit{decay mass} to distinguish it from the critical mass due to axion self-interactions.
Any AS whose mass is equal to or greater than the decay mass undergoes parametric resonance and, in turn, loses mass by emitting photons.
Meanwhile, an AS with a mass lighter than the decay mass accretes background axions, either from its host minicluster or from the NFW background, thus increasing its mass.
We hence expect that a large number of ASs pile up at the decay mass over time due to accretion.
We estimate the mass distribution, also known as the \textit{mass function}, of ASs in our galaxy and the number of ASs having the decay mass, following \cite{maseizik_pheno_2024}, which is based on the MC mass function, the core-halo relation \cite{Schive:2014dra,Schive:2014hza,veltmaat_formation_2018,eggemeier_formation_2019,chen_new_2021,Padilla:2020sjy,zagorac_soliton_2023, Kawai:2023okm}, and the accretion scenarios discussed in this work.

To calculate the AS mass function, we start with the MC mass function.
The MC mass function depends on the axion parameters, mass $m_a$ and decay constant $f_a$, while influenced by uncertainties or additional dependence, e.g., initial conditions of density perturbations, gravitational evolution, and tidal disruption \cite{kavanagh_stellar_2021,Dandoy:2022prp}.
All these factors boil down to a few parameters of the MC mass function: slope, amplitude, and both low- and high-mass cutoffs \cite{fairbairn_structure_2018, maseizik_pheno_2024}.
We use a slope $\alpha=-0.5$ based on the analytic approach of the Press-Schechter formalism \cite{Press_Schechter,fairbairn_structure_2018,maseizik_pheno_2024}.
The amplitude of the MC mass function is normalized by the total MC mass of the Milky Way, considering an MC abundance $f_{\rm MC} = 0.75$ based on numerical simulations in the post-inflationary scenario by \cite{eggemeier_first_2020}.
The binding fraction $f_{\rm MC}$ represents the fraction of axion dark matter bound in MCs, and $f_{\rm MC}=0.75$ indicates that 75\% of the axion dark matter is forming MCs.
We introduce the low- and high-mass cutoffs of the MC mass function in Sec.~\ref{sec:MCMF}.
By applying the core-halo relation by \cite{Schive:2014dra} to the MC mass function, we establish a connection between the MC mass function and the AS mass function.
We introduce several additional cutoffs for consistency of the AS-MC systems, as detailed in Sec.~\ref{sec:CoreHalo_ASMF} and \cite{maseizik_pheno_2024}.

We examine two accretion scenarios---the external and the internal accretion relative to the MC---in Secs.~\ref{sec:background-accretion-model} and \ref{sec:self-similar-model}, focusing on the AS-MC systems containing an AS at the decay mass and undergoing parametric resonance.
The accretion rate determines the photon spectral flux density for the parametric-resonant ASs.
The external accretion of ASs occurs in two steps.
First, the MC captures the axions from the galactic dark matter halo via gravitation, and a portion of the captured axions is thereafter transferred to the AS during the virialization of the AS-MC system.
The corresponding transfer rate can be calculated from the core-halo relation which applies to the virialized system.
On the other hand, the internal accretion supposes an evolution of the AS-MC system that deviates from the core-halo relation from \cite{Schive:2014dra} at late times.
According to numerical simulations in \cite{Dmitriev:2023ipv}, the AS grows by gravitational capturing of surrounding axion dark matter from its host MC.
The behavior of growth follows a so-called self-similar growth model \cite{Dmitriev:2023ipv}, which we use to determine the internal accretion rate for isolated AS-MC systems.

We present the resultant radio-line flux density and constraint plots in Sec.~\ref{sec:results}.
We analyze the signals using two methods: comparison with backgrounds and comparison with the sensitivities of telescopes such as LOFAR, FAST, ALMA, and SKA.
Backgrounds include the cosmic microwave background and the observed radio excess, parameterized as $T_{\rm bkg} = T_{\rm CMB} + 30.4\,{\rm K}\left(f/310\,{\rm MHz}\right)^{-2.58}$ \cite{Dowell:2018mdb}.
Interestingly, they yield constraints that appear mostly independent of the accretion scenarios, with $g_{a\gamma\gamma}\simeq 10^{-12} - 10^{-11}\,{\rm GeV}^{-1}$ for $m_a \simeq 10^{-7} - 10^{-2}\,{\rm eV}$.

The organization of the paper is as follows.
In Sec.~\ref{sec:ASMC}, we summarize the relevant features of ASs and MCs in our galaxy, including mass, radius, and their mass functions.
We briefly introduce the parametric resonance of ASs and their corresponding critical mass, the decay mass.
We then outline the MC and AS mass functions in our galaxy with the low- and high-mass cutoffs, which determine the number of ASs reaching the decay mass, as discussed in \cite{maseizik_pheno_2024}.
In Sec.~\ref{sec:accmodels}, we introduce the different accretion scenarios---external and internal accretion---for ASs as the solitonic cores of MCs.
AS-MC systems that are sufficiently massive for the AS core to reach the decay mass can emit photons in a narrow spectral line due to parametric resonance.
We estimate the resulting spectral flux density from the total number of decaying AS-MC systems observed at earth for different accretion models.
In Sec.~\ref{sec:results}, we summarize our constraints by comparing the predicted signals against their noise counterparts, including radio backgrounds and the instrumental noise of radio telescopes, parametrized by telescope sensitivities.
We infer constraints on the axion-photon coupling over a range of axion masses for each accretion model and discuss possible improvements of our work in the conclusion in Sec.~\ref{sec:discussion}.

\section{Axion Stars and Axion Miniclusters \label{sec:ASMC}}
\subsection{Axion star and parametric resonance\label{sec:AS-theory}}
The cosmological evolution of axion dark matter can be described by classical fields because of its large occupation numbers.
  Since the axion dark matter is generated from non-thermal origins such as the misalignment mechanism and the decay of topological defects, leading to the production of cold dark matter,
  the axion field can be approximated in the non-relativistic regime as \cite{hertzberg_merger_2020},
\begin{align}
    a(\mathbf{x},t) &=
     \frac{1}{\sqrt{2m_a}}\left[\psi(\mathbf{x}, t) e^{-i m_a t }+\psi^\ast(\mathbf{x}, t) e^{i m_a t}\right].
    \label{eq:axionnrel}
\end{align}
The equations of motion for self-gravitating axions in the non-relativistic regime with a Newtonian potential $\phi_N$ are called the Gross-Pitaevskii-Poisson (GPP) equations,
\begin{align}
\begin{split}\label{eq:Schrodinger}
    i\frac{\partial \psi}{\partial t}=-\frac{1}{2m_a}\nabla^2\psi+m_a\phi_N\psi-\frac{|\lambda|}{8m_a^2}|\psi|^2\psi,
\end{split}\\
\begin{split}
    \nabla^2\phi_N = 4\pi Gm_a|\psi|^2\label{eq:Poisson}
\end{split},
\end{align}
where $\lambda$ is the quartic coupling constant of the potential $V(a)=m_a^2 a^2/2 + \lambda a^4 / 4!$.
The stationary solutions of the GPP equations can be divided into two branches of solutions: the dilute stable branch and the dense unstable branch.
We focus on the dilute stable branch throughout this work because it is stable against perturbations.
In the dilute stable branch, gravitational interactions mostly dominate over self-interactions, whereas in the dense unstable branch, self-interactions dominate.

The critical AS configuration is the tipping point between the dilute stable branch and the dense unstable branch, where the AS has a maximum mass $M_{\star,\max}$.
Since the mass and radius are inversely proportional to each other on the dilute branch, the radius reaches its minimum $R_{\star,\min}$ at the critical AS configuration \cite{chavanis_mass-radius_2011,Visinelli:2017ooc,hertzberg_merger_2020}.
Using a Gaussian ansatz for the approximately stationary profile, $\psi(\mathbf{x},t) \simeq \psi(r)$,
 \begin{align}
 \psi(r)=\sqrt{\frac{M_\star}{\pi^{3/2}m_a R_\star^3}}\exp\left({-\frac{r^2}{2R_\star^2}}\right),
 \end{align}
one can find the critical AS configuration by extremizing the Hamiltonian of the system including the self-interaction and the gravitational interaction,
\begin{align}
    M_{\star,{\rm max}} &= 2.17\times 10^{-11} M_{\odot}\left(\frac{ {\rm \mu eV}}{m_a}\right)\left(\frac{f_a}{10^{11} \,{\rm GeV}}\right)\left(\frac{0.3}{\gamma}\right)^{1/2}, \label{eq:M_starmax}
    \\
    \label{eq:R_crit}
    R^{90}_{\star,{\rm min}} &= 4.02\times 10^3 \hspace{1mm}{\rm km}\left(\frac{ {\rm \mu eV}}{m_a}\right)\left(\frac{10^{11} \,{\rm GeV}}{f_a}\right)\left(\frac{\gamma}{0.3}\right)^{1/2},
\end{align}
where $R^{90}_{\star,{\rm min}}$ is defined as the radius enclosing 90\% of the AS mass, $R^{90}_{\star,{\rm min}}\simeq 1.768 R_{\star,{\rm min}}$, and $\gamma \equiv -\lambda f_a^2/m_a^2$ to parametrize the self-interaction $\lambda$.
The general mass and radius of the dilute stable ASs can be expressed in terms of the critical quantities in Eqs.~\eqref{eq:M_starmax} and \eqref{eq:R_crit} by introducing an additional parameter $\alpha\in(0,1]$ as
\begin{align}
    M_\star(R^{90}_\star)&=\alpha M_{\star,{\rm max}}(R^{90}_{\star,{\rm min}}), \label{eq:alpha}\\
    R^{90}_\star&=g(\alpha)R^{90}_{\star,{\rm min}}, \label{eq:g_alpha}
\end{align}
where $
    g(\alpha)=(1+\sqrt{1-\alpha^2})/{\alpha}$ \cite{hertzberg_merger_2020}.
    
 The axion-photon coupling in Eq.~\eqref{eq:axion-photon-Lagrangian} sets another critical mass for the AS, beyond which the AS decays through parametric resonance induced by ambient photons.\footnote{We consider the decay $a\rightarrow 2\gamma$ of axions into two photons within a soliton in our galaxy. The effective photon mass is given by $m_{\gamma}^2 \simeq 1.4\times 10^{-21}(X_e - 7.3\times 10^{-3}(\omega/{\rm eV})^2(1-X_e))(n_p/{\rm cm}^{-3})\,{\rm eV}^2$ \cite{1980PoO,Mirizzi:2009iz,Mirizzi:2009nq}, where $X_e = n_e/n_p$ is the ratio of the free electron density $n_e$ to the proton density $n_p$, and $\omega$ is the photon energy. In our galaxy, the free electron number density is $\mathcal{O}(10^{-2})\,{\rm cm}^{-3}$ in the galactic disk \cite{Cordes:2002wz}, and in the halo region of our interest, it would be significantly lower. The second term in the expression is negligible because the photon energy, being half the axion mass, is much less than eV. Therefore, the effective photon mass is smaller than $\mathcal{O}(10^{-12})\,{\rm eV}$. Since we are focused on axion masses ranging from $\mathcal{O}$(neV) to higher values, the effective photon mass is not relevant to our analysis.}
 To differentiate the critical AS mass resulting from the axion-photon coupling in \eqref{eq:axion-photon-Lagrangian} from the critical AS mass due to self-interaction $M_{\star,\max}$ in Eq.~\eqref{eq:M_starmax}, we refer to the former as the decay mass $M_{\star,\gamma}$.
 We obtain the decay mass by solving the equation of motion for photons \cite{Hertzberg:2018zte,levkov_radio-emission_2020,hertzberg_merger_2020}.
Parametric resonance occurs when the maximum growth rate inside the AS, calculated in the limit of the homogeneous condensate, $\mu_{\rm H} \simeq g_{a\gamma\gamma}m_a a_0/4$, becomes larger than the escape rate of the photon, $\mu_{\rm esc}\simeq (2R_\star)^{-1}$. Here, $a_0=\sqrt{2M_\star/(\pi^{3/2}m_a^2R_\star^3)}$ is the amplitude of the axion field at the center of the AS, i.e. at $r=0$.
  The growth rate of the parametric resonance can be approximated as $\mu_{\rm growth}\simeq \mu_{\rm H}-\mu_{\rm esc}$ \cite{Hertzberg:2018zte,hertzberg_merger_2020}.
  The escape rate of the photon encodes the profile of the AS, and we use the characteristic radius of the AS profile, $R_\star$, instead of $R_{\star}^{90}$ for estimating the escape rate.

For solitons with a Gaussian profile, the parametric resonance condition becomes \cite{chavanis_mass-radius_2011, Hertzberg:2018zte}
 \begin{align}
\label{Hertzbergconst1}
    g_{a\gamma\gamma}f_a > 0.23
    \left(\frac{\gamma}{0.3}\right)^{1/2}
    \left(\frac{g(\alpha)}{\alpha}\right)^{1/2}
    .
\end{align}
This resonance condition can be rephrased in terms of the AS mass $M_\star$ by expressing $\alpha$ and $g(\alpha)$ through Eqs.~\eqref{eq:alpha} and \eqref{eq:g_alpha}.
The resulting decay mass $M_{\star,\gamma}$, which is the critical AS mass induced by the axion-photon coupling, is
\begin{align}
    \label{eq:M_Decay_Gaussian}
    M_{\star,\gamma} &\simeq 1.6 \times 10^{-12} M_{\odot}\left(\frac{\mu{\rm eV}}{m_a}\right)\left(\frac{10^{-11} {\rm GeV}^{-1}}{g_{a\gamma\gamma}}\right)^2\left(\frac{10^{11} {\rm GeV}}{f_a}\right)
    \nonumber\\
    &\hspace{0.5cm}\times
    \sqrt{\left(\frac{g_{a\gamma\gamma}f_a}{0.23}\right)^2-\frac{5}{3} \left(\frac{\gamma}{1}\right)},
\end{align}
where the resonance occurs for ASs with $M_\star \geq M_{\star,\gamma}$.\footnote{The gravitational redshift could detune the resonance \cite{Arza:2020eik}. If we apply the resonant AS configuration to the condition that the resonance is not detuned by gravitational redshift, this condition is generally satisfied up to numerical factors of a similar order of magnitude after the complete cancellation of parameter dependence. Even if the numerical coefficients turn out not to satisfy the gravitational redshift condition, it does not mean that resonant ASs do not exist. Instead, the resonant AS configuration would be established with a slightly higher mass and smaller radius.}
ASs beyond this mass act as resonant amplifiers of ambient photons with frequencies $f \simeq m_a/(4\pi)$ through cascade-like stimulated emission.
We briefly outline how the existence of a decay mass $M_{\star,\gamma}< M_{\star,\max}$ can be combined with ongoing accretion onto the soliton to produce galactic background emission\footnote{See \cite{levkov_radio-emission_2020,Di:2023nnb} for the time evolution of the emission from collapsing ASs in the other limit, $M_{\star,\gamma}> M_{\star,\max}$.}:

If an AS is slightly heavier than the decay mass but lighter than the maximum stable AS mass, i.e. $M_{\star,\gamma} \lesssim M_\star < M_{\star,\max}$, the mass excess $\Delta M_\star \equiv M_\star - M_{\star,\gamma}$ gets dissipated into photons at half the frequency of the axion mass.
The resonant emission increases exponentially and shuts off when the AS reaches a subcritical state again, $M_\star < M_{\star,\gamma}$.
This relaxation process however competes with the continuous accretion of axion dark matter onto the AS (see Sec.~\ref{sec:accmodels}).
As a result of these two competing processes, we expect a pile-up of ASs around $M_\star \simeq M_{\star,\gamma}$, where the resonant solitons continuously emit photons through conversion of accreted axion dark matter $\Delta M_\star$.

Since the exponential growth of resonant emission can always compensate the linear mass accretion over time, a modulation of the radio signal may occur around the decay mass when the accretion rate is eventually overcompensated by the exponential conversion into radio photons.
For simplicity and due to lack of detailed knowledge at the time of publication, we assume that on average, parametric-resonant ASs emit the energy gained from accretion in a continuous way.
While the individual signals might modulate over time, the very large number of resonant systems involved motivates our simplified assumption of a continuous, averaged emission.

The key parameter for determining the strength of radio emission from parametric resonance of ASs is the total number of ASs at the decay mass $M_{\star,\gamma}$.
To determine the number of resonant AS-MC systems in our galaxy, and to develop appropriate accretion models for these systems in Sec.~\ref{sec:accmodels}, we need to infer the mass distribution of galactic MCs as well as that of their AS counterparts.
We use the parametrization of the MC mass distribution as given in \cite{fairbairn_structure_2018} to infer the AS mass distribution using the core-halo relation \cite{maseizik_pheno_2024} in the following subsections.

\subsection{Axion-minicluster mass function in our galaxy  \label{sec:MCMF}}

Let us start with the characteristic quantities of MCs, namely the mass, density, and radius for a given axion model.
Given that more substructures are formed in the post-inflationary scenario due to the larger initial density perturbations on small scales, we focus on the post-inflationary scenario.\footnote{Note that even in pre-inflationary scenarios, non-standard misalignment mechanisms---such as large misalignment, kinetic misalignment with fragmentation, and misalignment with non-periodic potentials---enhance the growth of density fluctuations, leading to even denser MCs \cite{Eroncel:2022efc,Chatrchyan:2023cmz}. This opens another interesting direction, associated with our accretion models.}
The abundance of axion dark matter in the post-inflationary scenario can be determined by the three axion parameters \cite{fairbairn_structure_2018}: the zero-temperature axion mass $m_{a,0}$, the axion decay constant $f_a$, and the temperature index $n$ setting the temperature evolution of the axion mass in the high-temperature regime, e.g., $T\gtrsim \mu \equiv \sqrt{m_{a,0}f_a}$,
\begin{align}
m_a(T > \mu)= m_{a,0} \left(\frac{T}{\mu} \right)^{-n}. \label{eq:m_T}
\end{align}
Although we consider $\mu=\sqrt{m_{a,0}f_a}$ as a benchmark, $\mu$ can vary from $\sqrt{m_{a,0}f_a}$ in principle. The case where $\mu\gg \sqrt{m_{a,0}f_a}$ is approximately equivalent to the $n=0$ case.
In the following sections, unless explicitly stated, we use $m_a\equiv m_{a,0}$ to denote the zero-temperature axion mass.
Given the dark matter abundance in the present universe, $\Omega_a h^2 = 0.12$ \cite{Planck:2018vyg}, only two of the three parameters ($m_a,f_a,n$) are independent.
We fix $f_a$ as a function of $m_a$ such that it satisfies the correct dark matter abundance $\Omega_a h^2 = 0.12$ for a given temperature index $n$, as in \cite{fairbairn_structure_2018, maseizik_pheno_2024} (see also the details in App.~\ref{app:DM_Basics}).

The characteristic mass of MCs, $\mathcal{M}_0$, can be estimated directly from the total mass of axion dark matter enclosed within a Hubble horizon at the oscillation temperature $T=T_\mathrm{osc}$, when the axion field starts to oscillate, $3H(T_\mathrm{osc}) \simeq m_a(T_\mathrm{osc})$, \cite{Kolb:1993zz,fairbairn_structure_2018}:
\begin{align}
    \mathcal{M}_0 = \rho_{a,0} \frac{4\pi}{3}\left(\frac{\pi}{a_{\rm osc} H_{\rm osc}}\right)^3
    .
    \label{eq:M0}
\end{align}
Here, $\rho_{a,0}$ denotes the axion dark matter density in the present universe.
$a_{\rm osc} = a(T_{\rm osc})$ and $H_{\rm osc} = H(T_{\rm osc})$ are the scale factor, normalized such that the scale factor in the present universe is $a_0 = 1$, and the Hubble parameter at the oscillation temperature, respectively.
Note that $\mathcal{M}_0$ is a rough estimate representing the mass distribution of the MCs.

The radius of the MC can be determined if the mass and density of the MC are specified.
The density of MCs depends on the initial overdensity parameter $\Phi = (\rho_a-\langle\rho_a\rangle) / \langle\rho_a\rangle|_{T_{\rm dec}<T<T_{\rm osc}}$, where the bracket represents the spatial average, and $T_{\rm dec}$ is the temperature when the MC decouples from cosmic expansion \cite{kolb_large-amplitude_1994,Tinyakov_2016}. The density is given by
\begin{align}
\rho_{\rm MC} &\simeq 7\times 10^6 \, \Phi^3 (1+\Phi)
\,\mathrm{GeV}/\mathrm{cm}^{3},
\label{eq:rho_mc}
\end{align}
which is calculated based on the spherical collapse model in \cite{kolb_large-amplitude_1994}.
Knowing the density of the MC, we can infer the radius of the MC with a given mass $\mathcal{M}$, assuming a spherically homogeneous MC profile for simplicity \cite{Tinyakov_2016}:
\begin{align}
\mathcal{R} &\simeq \frac{3.4\times 10^7}{\Phi(1 + \Phi)^{1/3}} \left(\frac{\mathcal{M}}{10^{-12}M_\odot}\right)^{1/3}\mathrm{km}. \label{eq:R_mc}
\end{align}

While the characteristic mass $\mathcal{M}_0$ in Eq.~\eqref{eq:M0} provides a reasonable estimate for a typical MC mass before the cosmological evolution around and after matter-radiation equality, the mass distribution of MCs evolves and significantly broadens over the cosmological evolution \cite{fairbairn_structure_2018,eggemeier_first_2020}.
This evolution of the mass distribution of MCs, referred to as the MC \textit{mass function} (MCMF), can be described by the Press-Schechter formalism.
The MCMF is defined as the number density distribution of MCs per unit volume per logarithmic mass interval, $(dn/d\ln \mathcal{M})(\mathbf{r})$. 
It can be parameterized in the simple form
\begin{align}
    \frac{dn}{d\ln \mathcal{M}}(\mathbf{r}) = C_{\rm n}(\mathbf{r})
    \left(
    \frac{ \mathcal{M}}{ \mathcal{M}_0}
    \right)^{\alpha}
    ,
\end{align}
where the slope $\alpha=-1/2$ is inferred from the analytic method based on the Press-Schechter formalism \cite{fairbairn_structure_2018}.
We normalize the MCMF using the normalization profile $C_{\rm n}(\mathbf{r})$, which is expected to follow the dark matter halo profile.
We integrate $C_{\rm n}(\mathbf{r})$ over the Milky Way volume and equate the total MC mass from the MCMF, $\mathcal{M}_\mathrm{tot}$, to the total DM mass bound in MCs within the NFW halo.
Throughout this work, we assume that about $75\%$ of the dark matter in our galaxy is bound to form MCs, based on the simulations in \cite{eggemeier_first_2020}:
\begin{align}
    \mathcal{M}_\mathrm{tot}
    &\simeq f_{\rm MC} M_{\rm MW},
    \label{eq:MMCtot}
    \\
    &\simeq \int_{\rm MW} d^3\mathbf{r}\int_{\mathcal{M}_{\min}}^{\mathcal{M}_{\max}}
    d\mathcal{M} 
    \frac{dn}{d\ln \mathcal{M}}(\mathbf{r}),
    \label{eq:dndlnMnorm}
\end{align}
where $f_{\rm MC}=0.75$ and $M_{\rm MW}$ is the total mass of dark matter within our galaxy. 
We parametrize the MCMF using the low-mass cutoff $\mathcal{M}_{0,\min}$ and the high-mass cutoff $\mathcal{M}_{0,\max}$, which depend on the evolution and initial mass of structures at different values of ($m_a,n$) as in \cite{fairbairn_structure_2018} and \cite{maseizik_pheno_2024}.
At present-day redshift $z=0$, these mass-cutoffs are given by \cite{fairbairn_structure_2018}
\begin{align}
\mathcal{M}_{0,\min}(m_a,n) \Big |_{z=0} &\simeq \mathcal{M}_0(m_a,n) / 25, \label{eq:M0_cutoff}
\\
\mathcal{M}_{0,\max}(m_a,n) \Big |_{z=0} &\simeq 4.9 \times 10^6 \mathcal{M}_0(m_a,n) \label{eq:M_h_max} .
\end{align}
Although the high-mass cutoff $\mathcal{M}_{0,\max}$ results from the continuous linear growth of structures with initial mass $\mathcal{M}_0$, the low-mass tail of the MCMF is subject to large uncertainties \cite{fairbairn_structure_2018}.
Because the normalization profile $C_{\rm n}(\mathbf{r})$ is not sensitive to the low-mass cutoff $\mathcal{M}_{0,\min}$, we can take a conservative approach by neglecting the low-mass tail of the MCMF with $\mathcal{M} \lesssim \mathcal{M}_0/25$.
One convenience for our signal is that the parametric-resonant AS-MC systems are primarily contributed by the high-mass component of the MCMF, which makes our results largely independent of any uncertainties in the low-mass cutoff, except when considering the parameter range of small resonant AS masses.

We repeat the above procedure for $n = 0,\, 1,\, 3.34$ over the range of axion masses $10^{-8}\,\text{eV} \leq m_a \leq 10^{-2}\,\text{eV}$.
The QCD axion corresponds to $m_a\simeq 50\,\mu$eV with $n=3.34$ \cite{Wantz:2009it}.
See an example plot for the MCMF in App.~\ref{app:ASMF}.

\subsection{Axion star mass function through core-halo relation\label{sec:CoreHalo_ASMF}}

In order to determine the abundance of parametric resonant AS-MC systems in our galaxy, we derive the AS mass function (ASMF) from the MCMF at $z=0$.
We assume that at most a single AS forms inside each MC, provided the AS inferred from the core-halo relation meets the consistency requirements for composite AS-MC systems, to be introduced in this section.
Because we are interested in the cosmologically stable configurations of the AS, i.e., the dilute stable branch of AS solutions, gravity is the dominant interaction over self-interaction. In this regime, the core-halo relation of pure gravity to virialize the AS-MC system can be applied \cite{Schive:2014dra,Schive:2014hza,eggemeier_formation_2019,chen_new_2021,maseizik_pheno_2024} to predict the core AS mass from its host MC mass:
\begin{align}
M_\star(z) & 
=  \mathcal{M}_{h,\min}(z) \left(\frac{\mathcal{M}}{\mathcal{M}_{h,\min}(z)}\right)^{1 / 3}  ,\label{eq:CoreHalo}
\end{align}
where the redshift-dependent minimum halo mass
\begin{align}
\mathcal{M}_{h,\min}(z) &= 8.34 \times 10^{-14}  M_\odot \, 
\left( \frac{1+z}{1+z_\mathrm{eq}} \right)^{3/4} \left(\frac{\zeta(z)}{\zeta(z_\mathrm{eq})}\right)^{1 / 4} \left(\frac{m_a}{\mu\text{eV}}\right)^{-3/2}, \label{eq:M_h_min_CoreHalo}
\end{align}
is defined by requiring $M_\star \geq \mathcal{M}$.
Here, $z_{\rm eq}$ is the redshift at matter-radiation equality, $z_{\rm eq}\simeq 3402$, and $\zeta(z)\equiv (18\pi^2 + 82 (\Omega_m(z)-1) - 39(\Omega_m(z)-1)^2)/\Omega_m(z)\simeq 18\pi^2$ for $z \gg 1$.
The redshift argument depends on the redshift of formation of the AS-MC system, where the overdense MC collapses gravitationally and thus decouples from the cosmic expansion.
Strictly speaking, this redshift depends on the overdensity parameter $\Phi$, but we approximate that AS formation occurs around $z\simeq z_{\rm eq}$ for simplicity.
Note that the redshift of AS formation is subject to open debate and that it generally depends on the MC properties as simulated in Refs.~\cite{Kolb:1993zz,levkov_gravitational_2018,eggemeier_formation_2019,chen_relaxation_2022,Kirkpatrick:2021wwz,Dmitriev:2023ipv}.

The ASMF can be obtained by applying Eq.~\eqref{eq:CoreHalo} to the MCMF in Sec.~\ref{sec:MCMF}.
However, additional cutoffs are required to ensure consistency for composite AS-MC systems. We implement these cutoffs in terms of the MC masses. Specifically, we introduce three additional low-mass cutoffs and one additional high-mass cutoff on top of the previous MCMF cutoffs in Eqs.~\eqref{eq:M0_cutoff} and \eqref{eq:M_h_max}. Let us begin with the low-mass cutoffs. 
1)~The first additional cutoff is given by Eq.~\eqref{eq:M_h_min_CoreHalo}, which ensures that the embedded AS core is not heavier than its host MC, i.e., $M_\star \leq \mathcal{M}$.
2)~Similarly, the second cutoff guarantees that the AS core is not larger in size than its host MC, $R_\star^{90} \leq \mathcal{R}$.
We use the core-halo relation \eqref{eq:CoreHalo} and mass-radius relations of ASs and MCs to find the radius-cutoff MC mass \cite{maseizik_pheno_2024},
\begin{align}
\mathcal{M}_{R,\min}(\Phi, z) 
&\simeq 4.38 \times 10^{-15}M_\odot \sqrt{\Phi^3 (1 + \Phi)}\, \left( \frac{1 + z_\mathrm{eq}}{1 + z} \right)^{3/4} \left( \frac{\zeta(z_\mathrm{eq})}{\zeta(z)} \right)^{1/4} \left(\frac{\mu\text{eV}}{m_a}\right)^{3/2},
\label{eq:M_h_min_RadiusCutoff}
\end{align}
where $z\simeq z_\mathrm{eq}$ again as a typical collapse redshift.
Eq.~\eqref{eq:M_h_min_RadiusCutoff} implies that for large $\Phi \gg 1$, the radius cutoff scales as $\mathcal{M}_{R,\min}\propto \Phi^2$ and thus yields the dominant low-mass cutoff for the densest MCs with $\Phi \in (0,10^4]$.
We checked the impact of $\mathcal{M}_{R,\min}$ and found that it does not affect our qualitative predictions, since the major contribution to the radio emission is given by AS-MC systems with $\Phi$ in the intermediate density range $1 \lesssim \Phi \lesssim 100$.
For convenience, let us define the resultant low-mass cutoff of the MCMF for consistent AS-MC systems as
\begin{align}
    \mathcal{M}_{\min}(\Phi) &\equiv \max[\mathcal{M}_{0,\min},\, \mathcal{M}_{R,\min}(\Phi),\, \mathcal{M}_{h,\min} ].
    \label{eq:MClowconsistent}
\end{align}
3)~Additional input for the low-mass cutoffs comes from the decay mass of ASs.
Since we consider only the parametric-resonant AS-MC systems, to calculate the total number of relevant MCs, we need to integrate the ASMF from the decay mass $M_{\star,\gamma}$ to the high-mass cutoff of the ASMF.
If we translate this condition into another cutoff, i.e., only consider $M_\star \geq M_{\star,\gamma}$, $M_{\star,\gamma}$ would act as a low-mass cutoff, or equivalently the corresponding MC mass $\mathcal{M}(M_{\star,\gamma})$ from the core-halo relation can be interpreted as a cutoff for the MCMF.
We define $\mathcal{M}_\gamma$ as the critical MC mass at which the MC is expected to contain a parametric-resonant AS in the present universe.
This way, $\mathcal{M}_\gamma$ is given by the core-halo relation $\mathcal{M}(M_{\star,\gamma})$ except when we consider the long-time evolution of AS due to accretion.
Taking long-time accretion into account, ASs with an initial mass smaller than the decay mass can still achieve resonance in the present universe through the long-time mass accumulation.
4)~Lastly, the additional high-mass cutoff arises from the maximum stable AS mass $M_{\star,\max}$, above which relativistic multi-particle interactions trigger axion bursts, also called bosenovae \cite{Eby:2016cnq,levkov_relativistic_2017}. 
In terms of the MC mass, we represent it as $\mathcal{M}(M_{\star,\max})$.
Combining all these additional cutoffs with the previous MCMF cutoffs in Eqs.~\eqref{eq:MClowconsistent} and \eqref{eq:M_h_max}, we obtain the \textit{combined low- and high-mass cutoffs},
\begin{align}
    \mathcal{M}_{\gamma,\min}(\Phi) &\equiv \max[\mathcal{M}_{\gamma}, \, \mathcal{M}_{\min}(\Phi)] \label{eq:M_h_min_total},
    \\
    \mathcal{M}_{\gamma,\max} &\equiv \min[\mathcal{M}(M_{\star,\max}),\, \mathcal{M}_{0,\max} ],
    \label{eq:M_h_max_total}
\end{align}
for parametric-resonant AS-MC systems.

However, there are some caveats to consider:
First, it requires further study to understand the evolution of parametric-resonant AS-MC systems when $\mathcal{M}_\gamma < \mathcal{M}_{\min}(\Phi)$.
In this case, we expect that the AS decay mass becomes smaller than any AS of the initial ASMF derived from the core-halo relation, and the host MC experiences significant back-reaction from the parametric-resonant core AS.
Additionally, this small $\mathcal{M}_\gamma$ regime corresponds to large $g_{a\gamma\gamma}$ values that are already constrained by helioscopes and astrophysical observations.
Therefore, we do not consider this regime in this work, except when discussing the long-time accretion of ASs.
In this context of long-time accretion, we allow configurations that deviate from the core-halo relation for the initial ASMF; in other words, we apply the core-halo relation in a non-strict manner for the $\mathcal{M}<\mathcal{M}_{\min}(\Phi)$ regime.
This leads directly to the next point.
Second, if we consider the long-time evolution of AS due to the accretion of axion dark matter, lower mass ASs below $M_{\star,\gamma}$ in the initial ASMF can grow to reach $M_{\star,\gamma}$ and contribute to the parametric resonance, i.e., lowering $\mathcal{M}_\gamma$ below $\mathcal{M}(M_{\star,\gamma})$.
We address this possibility in the following section when we discuss the accretion models, especially the internal accretion model where the accretion rate is higher than in the other models.
As mentioned right above, in this specific case, we widen the parameter range of our interest towards $\mathcal{M}(M_{\star,\gamma})<\mathcal{M}_{\min}(\Phi)$.
In short, we use the core-halo relation to estimate the initial ASMF from the MCMF under virialization only.
However, when considering the long-time accretion of ASs, we apply the core-halo relation to the initial ASMF more loosely, allowing for deviations from it, particularly in cases where $\mathcal{M}(M_{\star,\gamma})<\mathcal{M}_{\min}(\Phi)$.
For convenience, we summarize the parameters of MCs and ASs in Table~\ref{tab:Params}, located in App.~\ref{app:ASMF}.


\section{Accretion models for axion stars and estimated radio-line flux density\label{sec:accmodels}}

Axion dark matter should be organized into three hierarchical structures in our Milky Way galaxy: the dark matter halo, MCs, and ASs.
The dark matter halo establishes the overall gravitational potential of our galaxy, where approximately $75\%$ of dark matter may exist in the form of MCs \cite{eggemeier_first_2020}.
Based on numerical simulations and analytical understanding \cite{Kolb:1993zz,Schive:2014dra,Schive:2014hza,
fairbairn_structure_2018,
levkov_gravitational_2018,
veltmaat_formation_2018,
eggemeier_formation_2019,chen_new_2021,Dmitriev:2023ipv}, we conservatively assume that each MC contains one AS within it.
In Sec.~\ref{sec:CoreHalo_ASMF}, we derived the ASMF using the core-halo relation from the MCMF given in Sec.~\ref{sec:MCMF}, which is based upon the assumption of virialization.
In the presence of an axion-photon coupling, ASs with masses above the decay mass should convert their initial mass excess into radio photons, thus accumulating at the decay mass apart from virialization.
As discussed in Sec.~\ref{sec:AS-theory}, after this accumulation, the continuous accretion of axion dark matter onto these ASs at the decay mass leads to the ongoing conversion of axions into photons.

The accretion onto ASs can arise from two main sources: first from gravitational capturing of free axions from outside the MC and secondly by AS accretion from within the MC itself.
We term the first scenario as \textit{external accretion} and the second scenario as \textit{internal accretion}, which implies external/internal relative to the MC.
In the following subsections, we discuss these two AS accretion scenarios.
In each scenario, we calculate the mass-growth rate of the ASs.
The mass-growth rate of ASs at the decay mass is translated into the emission rate of radio photons at half the frequency of the axion mass.
These ASs at the decay mass are assumed to be continuously glowing within our galaxy.
Although their emission could potentially exhibit cyclic behavior, we approximate the average AS to emit continuously as mentioned before.

\subsection{External accretion\label{sec:background-accretion-model}}

Since according to the simulations in \cite{eggemeier_first_2020} roughly $25\%$ of the galactic DM should be gravitationally unbound, MCs can accrete background axion dark matter from the NFW halo of our galaxy.
As both the MCs and the background axions traverse the galaxy with typical velocities of $v\sim 239\,{\rm km/s}$, only a small fraction of background particles with relative velocities smaller than the MC's escape velocity can be captured.
Once the background axions are gravitationally bound in the MC, they can subsequently be transferred onto the AS.

\begin{figure}[t]
    \centering
    \includegraphics[width=\textwidth]{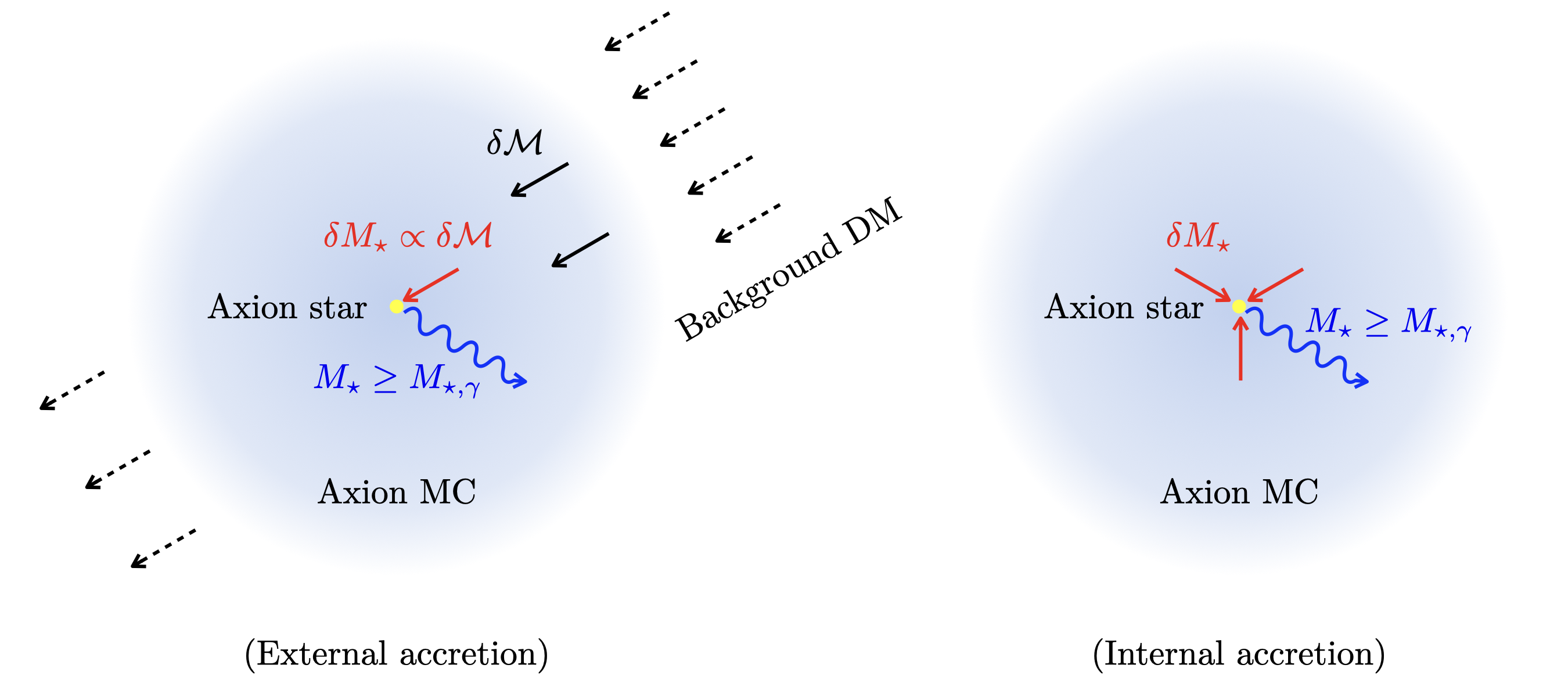}
    \caption{Schematic description of the external (left) and internal (right) accretion scenarios.
    In external accretion, the MC captures a small fraction of background axions whose relative velocities are smaller than the escape velocity of the MC through gravitational interaction.
    The black-dashed arrows represent the streams of background axions, while the black-solid arrows represent the fraction captured by the MC.
    The bound axions are then transferred further down to the AS, as depicted by the red arrow.
    The AS at the decay mass converts the mass excess into radio photons, as shown by the blue wavy arrow.
    In internal accretion, axions of the MC condense into the AS, where the AS accretes a fraction of the MC.
    Similarly, we denote the AS accretion with red arrows and the converted photons with a blue wavy arrow.}
    \label{fig:accretion_model}
\end{figure}

Fig.~\ref{fig:accretion_model} illustrates the external accretion scenario in the left panel.
Among the free streaming background dark matter within our galaxy, depicted with black-dashed arrows, only a fraction of axions have relative velocities small enough to be captured inside the MC.
This primary accretion onto the MC is depicted with black-solid arrows.
The resulting mass excess of axions captured by the MC is eventually virialized and partially accreted onto the AS, as indicated by the red arrow.
If the AS is at the decay mass, the accreted axions are converted into radio photons, as indicated by the blue wavy arrow.

We adopt the Navarro–Frenk–White (NFW) profile~\cite{navarro_structure_1996} for the background axion dark matter distribution in our galaxy,
\begin{align}
\label{eqn:NFWProfile}
    \rho_\mathrm{NFW}(r) \simeq \frac{0.25\, \rho_h}{\displaystyle \left(\frac{r}{r_h}\right)\left(1+\frac{r}{r_h}\right)^{2}}
    ,
\end{align}
where the numerical factor $0.25$ indicates the fraction of unbound dark matter~\cite{eggemeier_first_2020}, $\rho_h \simeq 0.32\,{\rm GeV}/{\rm cm}^3$, and $r_h\simeq 20.2\,{\rm kpc}$ \cite{mcmillan_mass_2011}.
We also performed our analysis using the Burkert profile for consistency, but found no significant impact.
Note that the distribution of MCs over our galaxy follows the profile of the background dark matter, i.e., $dn/d\ln\mathcal{M}\propto \rho_{\rm NFW}(r)$ according to Eq.~\eqref{eq:dndlnMnorm} and that we treat $\mathcal{M}, \Phi$ and the galactocentric coordinate $r$ as independent variables.

We estimate the geometrical capture rate of background axion dark matter by an MC with mass $\mathcal{M}$, located at radial position $r$ in our galaxy, as
\begin{align}
\label{eq:deltaM_mc}
    \frac{\delta \mathcal{M}}{ \delta t}(\Phi, \mathcal{M}, r)
    &=4\pi\mathcal{R}(\Phi,\mathcal{M})^2 \rho_{\rm NFW}(r)\langle v_{\rm cap}(\Phi,\mathcal{M},r)\rangle,
\end{align}
where $\mathcal{R}(\Phi, \mathcal{M})$ is the radius of the MC, given in Eq.~\eqref{eq:R_mc}, and the average relative velocity of the captured particles is
\begin{align}
     \label{vavg}
     \langle v_{\rm cap}(\Phi,\mathcal{M},r)\rangle
     &=\int_0^{v_{\mathrm{esc}}(\Phi,\mathcal{M})} dv \, v \, f_{v}(r),     \\
     &=\frac{v_{\rm vir}(r)}{\sqrt{2\pi}}
     \left[1-\exp\left(-\frac{v_{\mathrm{esc}}(\Phi,\mathcal{M})^2}{2v_{\rm vir}(r)^2}\right)
     \right],
     \\
     &
     \approx
     \frac{1}{2\sqrt{2\pi}}
     \frac{v_{\mathrm{esc}}(\Phi,\mathcal{M})^2}{v_{\rm vir}(r)}
     .
\end{align}
Here, we assume a one-dimensional Gaussian distribution for the relative velocity distribution, 
\begin{align}
f_v(r) &= \frac{1}{\sqrt{2\pi}\,v_{\rm vir}(r)} \exp{\left(-\frac{v^2}{2v_{\rm vir}(r)^2}\right)}
,
\qquad
v\in (-\infty, \infty),
\end{align}
at the galactocentric radius $r$, with the virial velocity $v_{\rm vir}(r)\simeq 239\,{\rm km/s}$ \cite{mcmillan_mass_2011} of the Milky Way rotation curve as the velocity dispersion.
We approximate $v_{\rm vir}(r)\simeq 239\,{\rm km/s}$ \cite{mcmillan_mass_2011,Irrgang:2012fw,Bhattacharjee:2013exa,mcmillan_mass_2017} as a constant throughout this work, which is valid over the relevant radial range from 1\,kpc to $\mathcal{O}(100)\,{\rm kpc}$.

To obtain the secondary AS mass growth rate, $\delta M_\star/\delta t$, from the MC mass growth rate, $\delta \mathcal{M}/\delta t$, given by Eq.~\eqref{eq:deltaM_mc}, we use the core-halo relation in Eq.~\eqref{eq:CoreHalo}.
This estimate is conservative because the resonant AS-MC system would have equal or smaller AS masses than the corresponding core mass, and the system tends to evolve towards virialization, which boosts the AS mass growth rate.
We furthermore neglect the direct capture of background axion dark matter by the AS core because of its smaller size and mass compared to the MC for simplicity.
This calculation is implicitly based on the observation that the fraction of accreted dark matter in the MC is very small even after the Hubble time, $(\delta \mathcal{M} / \delta t)\,t_H \ll \mathcal{M}$.
By differentiating the core-halo relation with respect to time, we can express the mass growth rate of the AS in terms of the mass growth rate of the MC:
\begin{align}
\label{eq:minimal}
    \frac{\delta M_\star^{\it ext}}{ \delta t}(\Phi, \mathcal{M}, r)
    &=
    \frac{1}{3}\left(\frac{\mathcal{M}}{\mathcal{M}_{h,\min}}\right)^{-\frac{2}{3}}\frac{\delta \mathcal{M}}{ \delta t}(\Phi, \mathcal{M}, r),
    \\
    &=\frac{4\pi}{3}\left(\frac{\mathcal{M}}{\mathcal{M}_{h,\min}}\right)^{-\frac{2}{3}} \mathcal{R}(\Phi,\mathcal{M})^2 \rho_{\rm NFW}(r)\langle v_{\rm cap}(\Phi,\mathcal{M},r)\rangle,
    \\
    &\simeq\frac{\sqrt{2\pi}}{3}\left(\frac{\mathcal{M}}{\mathcal{M}_{h,\min}}\right)^{-\frac{2}{3}} \mathcal{R}(\Phi,\mathcal{M})^2 \rho_{\rm NFW}(r)
    \frac{v_{\mathrm{esc}}(\Phi,\mathcal{M})^2}{v_{\rm vir}(r)}
    .
    \label{eq:dMdtext}
\end{align}
Once the AS reaches the decay mass, $M_{\star, \gamma}$, any mass excess $\delta M_\star = M_\star - M_{\star, \gamma}$ is converted into photons with a frequency $f \simeq m_a/(4\pi)$.
This process is indicated by the blue wavy line in Fig.~\ref{fig:accretion_model}.
We can calculate the flux density resulting from the conversion by dividing the mass growth rate by the surface area of a sphere with a radius equal to the distance between the source and the detection.

The flux density received from a single AS depends on the initial overdensity parameter $\Phi= \delta\rho_a/\bar{\rho}_a|_{t=t_i}$, the host MC mass $\mathcal{M}$, and the AS position $\mathbf{r}$ through the equation,
\begin{align}
    F_{\star,{\rm single}} &=
    \eta_{a\gamma}\frac{c^2}{4 \pi d_{\rm obs}(\mathbf{r})^2} \frac{\delta M_\star}{\delta t}(\Phi,\mathcal{M},r),
    \label{eq:Fstarsingle}
\end{align}
where $d_{\rm obs}(\mathbf{r})$ is the distance from the AS to the earth, $\eta_{a\gamma}$ denotes the efficiency with which the mass excess of ASs at the decay mass converts into photons, taken to be 1 in our work, and $c$ is the speed of light.
Eq.~\eqref{eq:Fstarsingle} is general, and the $\Phi$ dependence appears through the mass growth rate $\delta M_\star/\delta t$ and potentially the efficiency $\eta_{a\gamma}$.
This efficiency factor could even be time-dependent if the resonant emission occurs with cyclic features around the steady configuration.
As we assume a steady state for the accreting resonant AS, $\eta_{a\gamma}$ should be constant over time, especially unity, which implies the complete conversion of the mass excess beyond the decay mass into photons. If not, the mass excess of the AS would pile up and modulate the emission.

We can calculate the total flux density by multiplying $F_{\star,{\rm single}}$ by the total number of parametric-resonant ASs in our galaxy, $N^{\rm res}_\star$,
\begin{align}
F_{\star,\mathrm{tot}}
\simeq N^{\rm res}_\star F_{\star,{\rm single}}.
\nonumber
\end{align}
However, $F_{\star,{\rm single}}$ depends on $\Phi$, $\mathcal{M}$, and $\mathbf{r}$, and $N^{\rm res}_\star$ is obtained by integrating the number of ASs over those parameters.
Since the ASMF can be related to the MCMF via the core-halo relation, we can express the total flux density in terms of the MC distributions over $\mathcal{M}$ and $\Phi$.
The total flux density is then generally given by
\begin{align}
    F_{\star,\mathrm{tot}} &\simeq 
    \int_{{\rm MW},\,r>R_{\rm surv}} d^3\mathbf{r}
    \int_0^{10^4} d\Phi\,p_\Phi(\Phi)  \int_{\mathcal{M}_{\gamma,\min}(\Phi)}^{\mathcal{M}_{\gamma,\max}}
    d\mathcal{M} \, \frac{dn}{d\mathcal{M}}(\mathbf{r})\,
    F_{\star,{\rm single}},
    \label{eq:Ftotgeneral1st}
    \\
     &=\eta_{a\gamma} c^2 
    \int_0^{10^4} d\Phi\,p_\Phi(\Phi)  \int_{\mathcal{M}_{\gamma,\min}(\Phi)}^{\mathcal{M}_{\gamma,\max}}
    d\mathcal{M}
    \int_{{\rm MW},\,r>R_{\rm surv}}
    d^3\mathbf{r} \, \frac{1}{4 \pi d_\mathrm{obs}(\mathbf{r})^2} 
    \frac{dn}{d\mathcal{M}}(\mathbf{r})\,\frac{\delta M_\star}{\delta t},
    \\
     &\simeq 
     \eta_{a\gamma} c^2 
    \int_0^{10^4} d\Phi\,p_\Phi(\Phi)     \int_{\mathcal{M}_{\gamma,\min}(\Phi)}^{\mathcal{M}_{\gamma,\max}}
    d\mathcal{M}
    \int_{R_{\rm surv}}^{R_{200}} dr \, \left(\frac{r}{{\rm max}(r, R_\odot)}\right)^2 \frac{dn}{d\mathcal{M}}(r)\,\frac{\delta M_\star}{\delta t}, 
    \label{eq:F_total_general}
\end{align}
which using Eq.~\eqref{eq:dMdtext} for the external accretion rate gives,
\begin{align}
    F_{\star,\mathrm{tot}}^{\textit{ext}}
    &\simeq 
    \frac{ \sqrt{2\pi} }{3} \eta_{a\gamma}c^2
    \int_0^{10^4} d\Phi\,p_\Phi(\Phi)
    \int_{\mathcal{M}_{\gamma,\min}(\Phi)}^{\mathcal{M}_{\gamma,\max}}
    d\mathcal{M} \left(\frac{ \mathcal{M} }{\mathcal{M}_{h,\min}}\right)^{-\frac{2}{3}}\,
    \nonumber\\
    & \hspace{2cm} \times \mathcal{R}(\Phi, \mathcal{M})^2 \frac{v_\mathrm{esc}(\Phi,\mathcal{M})^2}{v_{\rm vir}}\int_{R_{\rm surv}}^{R_{200}} dr \, \left(\frac{r}{{\rm max}(r, R_\odot)}\right)^2 \rho_{\rm NFW}(r)
    \frac{dn}{d\mathcal{M}}(r), 
    \\
    &\simeq 
    \frac{ \sqrt{2\pi} }{3} \eta_{a\gamma}c^2
    \int_0^{10^4} d\Phi\,p_{\Phi,{\rm init}}(\Phi) \mathcal{P}_{\rm surv}(\Phi)
    \int_{\mathcal{M}_{\gamma,\min}(\Phi)}^{\mathcal{M}_{\gamma,\max}}
    d\mathcal{M} \left(\frac{ \mathcal{M} }{\mathcal{M}_{h,\min}}\right)^{-\frac{2}{3}}\,
    \nonumber\\
    & \hspace{2cm} \times \mathcal{R}(\Phi, \mathcal{M})^2 \frac{v_\mathrm{esc}(\Phi,\mathcal{M})^2}{v_{\rm vir}}\int_{R_{\rm surv}}^{R_{200}} dr \, \left(\frac{r}{{\rm max}(r, R_\odot)}\right)^2 \rho_{\rm NFW}(r)
    \frac{dn}{d\mathcal{M}}(r).
    \label{eq:F_totalext}
\end{align}
Note that we integrate over the volume of the Milky Way, starting from the survival radius $R_{\rm surv}\simeq 1\,{\rm kpc}$, within which MCs are unlikely to survive due to tidal disruption in the galactic bulge \cite{kavanagh_stellar_2021}, and up to the virial radius $R_{200}\simeq \mathcal{O}(100)\,{\rm kpc}$.
The contribution to radio signals from the outskirts beyond 100\,kpc will be relatively insignificant due to the low density of dark matter and the large distances to the earth.
Here, $p_\Phi(\Phi)$ denotes the probability distribution of the overdensity parameter $\Phi$, which results from cosmic evolution including tidal disruption.
To include the effects of tidal disruption, we introduce the survival probability $\mathcal{P}_{\rm surv}(\Phi)$ \cite{Dandoy:2022prp}, that is multiplied by the initial probability distribution $p_{\Phi,{\rm init}}(\Phi)$ (see App.~\ref{app:densitypar}).
We simplify the spatial dependence by approximating $d_{\rm obs}(\mathbf{r}) = |\mathbf{r}-\mathbf{R}_E| \simeq {\rm max}(r,R_\odot)$, based on the average observation distances $\langle d_{\rm obs} (\mathbf{r})\rangle_{r\leq R_\odot}\simeq R_\odot$ and $\langle d_{\rm obs} (\mathbf{r})\rangle_{R_\odot<r\leq \rho}\simeq \rho$.
Here, the brackets represent the spatial average over the indicated volume in the subscripts.
Additionally, the virial velocity is approximated as a constant $v_{\rm vir} \simeq 239\,{\rm km/s}$ as before.

Note that the mass accretion rate of the AS in Eq.~\eqref{eq:minimal} includes a suppression factor $\left( \mathcal{M} /\mathcal{M}_{h,\min}\right)^{-2/3}\ll 1$, which originates from the core-halo relation.
However, the parametric-resonant AS-MC system can deviate from the core-halo relation due to the deficit in AS mass caused by the resonance.
The system tends to evolve into a virialized state, which may result in an enhanced accretion rate of ASs, as observed in numerical simulations \cite{Dmitriev:2023ipv}.
In this regard, the estimation in Eq.~\eqref{eq:minimal} is conservative and gives a minimal value.
We include the possibility of an enhanced accretion rate of the AS by simply omitting the suppression factor $\left( \mathcal{M} /\mathcal{M}_{h,\min}\right)^{-2/3}$ for comparison.
This approach constitutes another accretion rate in the external accretion scenario, which we refer to as the \textit{maximal external accretion} model in the following.
The maximum external accretion model can thus be seen as a less conservative estimate of the expected virialization rate of the AS-MC system, for which a more detailed numerical analysis beyond the scope of our work would be needed.

Since we consider that the mass input originates from the background axion dark matter, the mass growth rate of the AS should be smaller than that of the host MC; otherwise, the mass of the host MC would decrease.
Likewise in Eq.~\eqref{eq:F_totalext}, the total flux density for the \textit{maximal external accretion} model becomes
\begin{align}
    F_{\star,\mathrm{tot}}^{\textit{ext},\,\max}
    &= \frac{ \sqrt{2\pi} }{3}  \eta_{a\gamma}c^2
    \int_0^{10^4} d\Phi\,  p_{\Phi,{\rm init}}(\Phi) \mathcal{P}_{\rm surv}(\Phi) \int_{\mathcal{M}_{\gamma,\min}(\Phi)}^{\mathcal{M}_{\gamma,\max}} d\mathcal{M} \, \mathcal{R}(\Phi, \mathcal{M})^2 \frac{v_\mathrm{esc}(\Phi,\mathcal{M})^2}{v_{\rm vir}}
    \nonumber\\
    & \hspace{3.5cm} \times   \int_{R_{\rm surv}}^{R_{200}} \, dr \,\left(\frac{r}{{\rm max}(r, R_\odot)}\right)^2 \rho_{\rm NFW}(r)\, \frac{dn}{d\mathcal{M}}(r)
    .
    \label{eq:F_total_high}
\end{align}

\subsection{Internal accretion\label{sec:self-similar-model}}

Recent numerical simulations \cite{Dmitriev:2023ipv} suggest that even in isolated AS-MC systems, i.e. without MC mass accretion from the NFW background, the AS core can continuously deprive its host MC of part of its mass over time (see \cite{levkov_gravitational_2018,eggemeier_formation_2019,Dmitriev:2023ipv} for previous simulations on the isolated AS-MC systems).
We refer to this mass transfer from an isolated MC onto the AS core as \textit{internal accretion} and derive the mass growth rate of the AS based on the AS-MC evolution model in \cite{Dmitriev:2023ipv}.
One subtle point in this internal accretion scenario is that, according to the numerical simulations \cite{Dmitriev:2023ipv}, the core-halo relation does not correspond to a stationary configuration, as we assume in the external accretion.
Although the mass growth of ASs slows down after reaching the point where the core-halo relation is satisfied, the AS mass continues to increase afterwards, as observed in the simulations.
Therefore, when deriving the initial ASMF to estimate accretion rates, we take two approaches.
First, as in the external accretion, we suppose the core-halo relation as a strict condition between the MCMF and the initial ASMF.
If the initial ASMF cannot be derived from the core-halo relation due to a small decay mass, or equivalently, if in terms of MC masses $\mathcal{M}_\gamma<\mathcal{M}_{\min}(\Phi)$, we exclude this parameter regime from the calculation of the mass accretion rate.
Secondly, in light of the numerical simulations, we allow deviations of the initial ASMF from the core-halo relation and take into account the long-time accretion of ASs.
In other words, we also explore the parameter range where $\mathcal{M}(M_{\star,\gamma})<\mathcal{M}_{\min}(\Phi)$ and consider the long-time evolution of ASs.
Based on modelling the numerical simulations, after deriving the initial ASMF, we can predict how the initial ASs evolve over the long-time accretion within the MCs.

The evolution of the AS inside the MC can be described using a kinetic theory, resulting in the self-similar growth model.
The resulting mass transfer can be predicted by a simple fitting formula based on the simulations in \cite{Dmitriev:2023ipv}, which describes the relation between the masses of AS and MC over time:
\begin{align}
    \left(1 + \frac{1}{\epsilon^2 }\left(\frac{M_\star}{\mathcal{M}}\right)^3 \right)^3 \left(1 - \frac{M_\star}{\mathcal{M}} \right)^{-5} \approx \frac{t/\tau_{\rm gr} + 0.1}{1.1} 
    ,
    \label{eq:accretion_Levkov}
\end{align}
where\footnote{The explicit parameter dependence of $\epsilon$ is presented in the second arXiv version of \cite{Dmitriev:2023ipv}.}
\begin{align}
    \epsilon \simeq 0.086 \sqrt{\Phi} (1+\Phi)^{1/6} \left( \frac{10^{-12}\, M_\odot}{\mathcal{M}} \right)^{2/3} \left( \frac{\mu\text{eV}}{m_a} \right)
    ,
    \label{eq:epsilon}
\end{align}
and the kinetic time, also called the condensation time, is \cite{levkov_gravitational_2018}
\begin{align}
\tau_{\rm gr} \simeq \frac{5.7\times 10^6 \,\mathrm{yr}}{\Phi^3(1+\Phi)}\left(\frac{\mathcal{M}}{10^{-12} M_{\odot}}\right)^2\left(\frac{m_a}{\mu \mathrm{eV}}\right)^3
.
\label{eq:condtime}
\end{align}
In Eq.~\eqref{eq:accretion_Levkov}, the initial condition is set by $M_\star = 0$ at $t=\tau_{\rm gr}$, and we neglect the small contribution from excited states around the AS.

We derive the AS mass growth rate by taking the time derivative of Eq.~\eqref{eq:accretion_Levkov}.
Since this growth rate refers to the mass transfer from the MC to the AS, which is an internal process within the AS-MC system, we call this scenario \textit{internal accretion}.
The resulting mass growth rate in this scenario is given by
\begin{align}
    \frac{\delta M_\star^{\it int}}{\delta t}
    \simeq
    \frac{ \displaystyle
    \left(1-\frac{M_\star}{\mathcal{M}}\right)^6}
    {\displaystyle
    \left[5+\frac{1}{\epsilon^2}\left(\frac{M_\star}{\mathcal{M}}\right)^2\left(9-4\frac{M_\star}{\mathcal{M}}
    \right)\right]\left[1+\frac{1}{\epsilon^2}\left(\frac{M_\star}{\mathcal{M}}\right)^3\right]^2}
    \frac{\mathcal{M}}{1.1\,\tau_{\rm gr}}
    \label{eq:genaccrate}
    .
\end{align}
The AS mass growth \eqref{eq:genaccrate} amounts to a function of the core mass $M_\star$ and the MC mass $\mathcal{M}$.
Note that it implicitly depends on the overdensity parameter $\Phi$ and the axion mass $m_a$ as well.
As we are interested in AS-MC systems undergoing parametric resonance, the accretion rate can be estimated by taking $M_\star = M_{\star,\gamma}$ for a given $\mathcal{M}$ in Eq.~\eqref{eq:genaccrate}.

The total flux density can then be obtained by integrating the flux density from a single AS over the overdensity parameter $\Phi$, the MC mass $\mathcal{M}$, and the spatial MW volume as in Eqs.~\eqref{eq:Ftotgeneral1st} - \eqref{eq:F_total_general}:
\begin{align}
    F_{\star,\mathrm{tot}}^{\it \,int} 
      &\simeq 
     \eta_{a\gamma} c^2 
    \int_0^{10^4} d\Phi\,p_\Phi(\Phi)    \int_{\mathcal{M}_{\gamma,\min}(\Phi)}^{\mathcal{M}_{\gamma,\max}}
    d\mathcal{M}
    \int_{R_{\rm surv}}^{R_{200}} dr \, \left(\frac{r}{{\rm max}(r, R_\odot)}\right)^2 
    \nonumber\\
    &\hspace{2.5cm}\times \frac{dn}{d\mathcal{M}}(r)
    \,\frac{\delta M_\star^{\it int}}{\delta t}\bigg|_{M_\star = M_{\star,\gamma}}
    \Theta \left(\mathcal{M} - \frac{\delta M_\star^{\it int}}{\delta t}\bigg|_{M_\star = M_{\star,\gamma}}t_H  - \mathcal{M}_{\gamma} \right),
    \\
    &\simeq 
     \eta_{a\gamma} c^2 
    \int_0^{10^4} d\Phi\,p_{\Phi,{\rm init}}(\Phi)   \mathcal{P}_{\rm surv}(\Phi)   \int_{\mathcal{M}_{\gamma,\min}(\Phi)}^{\mathcal{M}_{\gamma,\max}}
    d\mathcal{M}
    \int_{R_{\rm surv}}^{R_{200}} dr \, \left(\frac{r}{{\rm max}(r, R_\odot)}\right)^2 
    \nonumber\\
    &\hspace{2.5cm}\times \frac{dn}{d\mathcal{M}}(r)
    \,\frac{\delta M_\star^{\it int}}{\delta t}\bigg|_{M_\star = M_{\star,\gamma}}
    \Theta \left(\mathcal{M} - \frac{\delta M_\star^{\it int}}{\delta t}\bigg|_{M_\star = M_{\star,\gamma}}t_H  - \mathcal{M}_{\gamma} \right),
    \label{eq:F_tot_new}
\end{align}
where we additionally introduced a Heaviside function, which imposes the condition that the host MC is not engulfed by the parametric-resonant AS core over the Hubble time. Specifically, the MC mass should be equal to or larger than the value given by the core-halo relation for the resonant AS at the decay mass, $\mathcal{M}_\gamma = \mathcal{M}(M_{\star,\gamma})$, plus the mass accreted by the AS during one Hubble time.
We apply this condition because, in the internal accretion scenario, the mass growth rate can grow significantly larger than that in the external accretion scenario, especially for large $\Phi$.
The back-reaction to the MC mass thus needs to be taken into account in this scenario.
As before, we also consider the effect of tidal disruption by including the survival probability $\mathcal{P}_{\rm surv}(\Phi)$ (see details in App.~\ref{app:densitypar}).

We make another attempt at exploring the possibilities of long-time AS accretion observed in \cite{Dmitriev:2023ipv}.
In the external accretion models and the first approach to internal accretion represented by Eq.~\eqref{eq:F_tot_new}, we take the initial ASMF adhering to the core-halo relation and implicitly neglect the mass growth of ASs beyond the initial AS mass, while excluding cases where $\mathcal{M}_\gamma<\mathcal{M}_{\min}(\Phi)$.
However, in the second approach, we allow the initial ASMF to stray from the core-halo relation and, consequently, consider the long-time AS accretion, as predicted in the simulations of \cite{Dmitriev:2023ipv}, i.e., by the corresponding self-similar growth model \eqref{eq:accretion_Levkov}.
This approach encompasses the parameter regime of $\mathcal{M}(M_{\star,\gamma})<\mathcal{M}_{\min}(\Phi)$.

As a simplified approach, we explore how long-time AS accretion can boost the number of resonant AS-MC systems.
This long-time AS accretion extends the $\mathcal{M}$-range of potentially resonant AS-MC systems in the present universe to lower MC masses than the decay mass $\mathcal{M}_{\gamma}=\mathcal{M}(M_{\star,\gamma})$ obtained from the core-halo relation.
We implement this simplified treatment by shifting the corresponding MC mass $\mathcal{M}_\gamma$.
The numerical simulations suggest that an order-one fraction of the MC mass could be accreted by the AS over long timescales, $t\sim t_H$.
We introduce the factor $C_\mathrm{acc}$ to account for this \textit{long-time} AS accretion and replace the low-mass cutoff $\mathcal{M}_\gamma$ in Eq.~\eqref{eq:M_h_min_total} obtained from the core-halo relation \eqref{eq:CoreHalo} by
\begin{align}
    \mathcal{M}_\gamma &= C_\mathrm{acc} M_{\star,\gamma},
    \label{eq:MMCgammalong}
\end{align}
where we set $C_\mathrm{acc}\simeq 2$, assuming that the AS core absorbs up to $50\%$ of the host MC mass.
In this simplified model, we use the observation of \cite{Dmitriev:2023ipv}, showing that the ASs absorb up to $\mathcal{O}(10)\%$ of their host MCs over long time scales, which is summarized in Eq.~\eqref{eq:accretion_Levkov}.
In the limit of $t \gg \tau_{\rm gr}$, the authors of \cite{Dmitriev:2023ipv} suggested that $M_\star \rightarrow \mathcal{M}$, where the AS core can even contain an order one fraction of the total AS-MC mass.
In general, $C_\mathrm{acc}$ is a complicated function depending on the initial MC mass; its mass evolution over time due to successive MC mergers, tidal disruption, and background accretion; as well as the overdensity parameter $\Phi$.
The total flux density has the same form as in Eq.~\eqref{eq:F_tot_new}, except with $\mathcal{M}_\gamma = \mathcal{M}(M_{\star,\gamma})$ replaced by $\mathcal{M}_\gamma = C_\mathrm{acc} M_{\star,\gamma}$ as in Eq.~\eqref{eq:MMCgammalong}:
\begin{align}
    F_{\star,{\rm tot}}^{{\it int},{\rm long}} \simeq F_{\star,{\rm tot}}^{\it int}(\mathcal{M}_\gamma = C_\mathrm{acc} M_{\star,\gamma})
    .
    \label{eq:Ftotintlong}
\end{align}

Note that $\tau_{\rm gr}$ in Eq.~\eqref{eq:condtime} can become larger than the Hubble time, depending on the parameters.
In such cases, an order-one fraction of the MC mass will not be absorbed by the AS within a Hubble time, which leads to $C_{\rm acc}>2$ in Eq.~\eqref{eq:MMCgammalong}.
However, the condensation time $\tau_{\rm gr}$ is derived under the assumption of a homogeneous background density field or a smoothly varying MC density profile \cite{levkov_gravitational_2018}, such as the NFW profile.
Here, smoothness is defined by comparing the scales with the axion field coherence length $(m_a v)^{-1}$, which is relevant to AS dynamics.
This condition for applying the condensation time does not hold during the early evolution of AS-MC systems, referred to as the violent relaxation phase \cite{eggemeier_formation_2019}, where virialization has not been completed.
This phase can be induced by MC evolution in the early universe---such as mergers, tidal disruption, and accretion---leading to the time-dependent gravitational potential and amplified density fluctuations.
In this early phase, the mass growth of the AS could be significantly enhanced compared to the typical condensation time, as mentioned in \cite{eggemeier_formation_2019}.
Due to this enhancement in AS formation, throughout this work, we assume that the initial formation of the AS core occurs quickly enough and examine the long-time evolution effect using the simplified approach with the constant $C_{\rm acc} \simeq 2$.

\section{Radio-line signals from accreting AS-MC systems\label{sec:results}}

\subsection{Expected spectral flux density\label{sec:flux-rates}}

\begin{figure}[t]
    \centering
    \begin{subfigure}{\textwidth}
    \includegraphics[width=\linewidth]{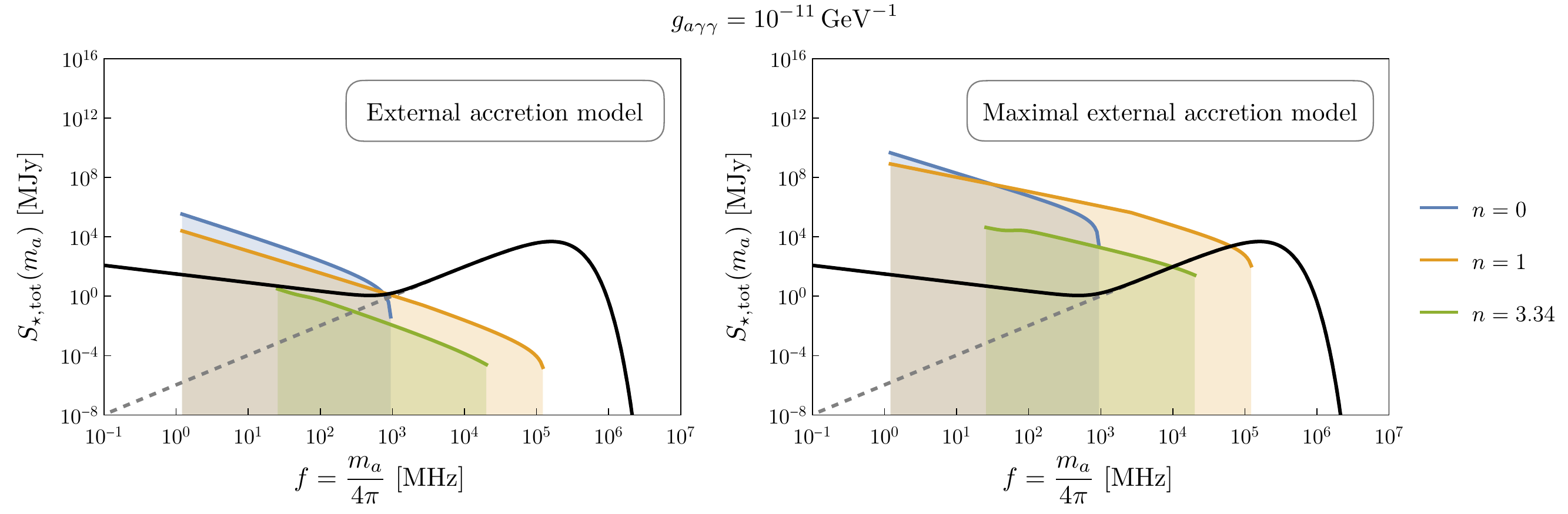}
    \end{subfigure}
    \begin{subfigure}{\textwidth}
    \includegraphics[width=\linewidth]{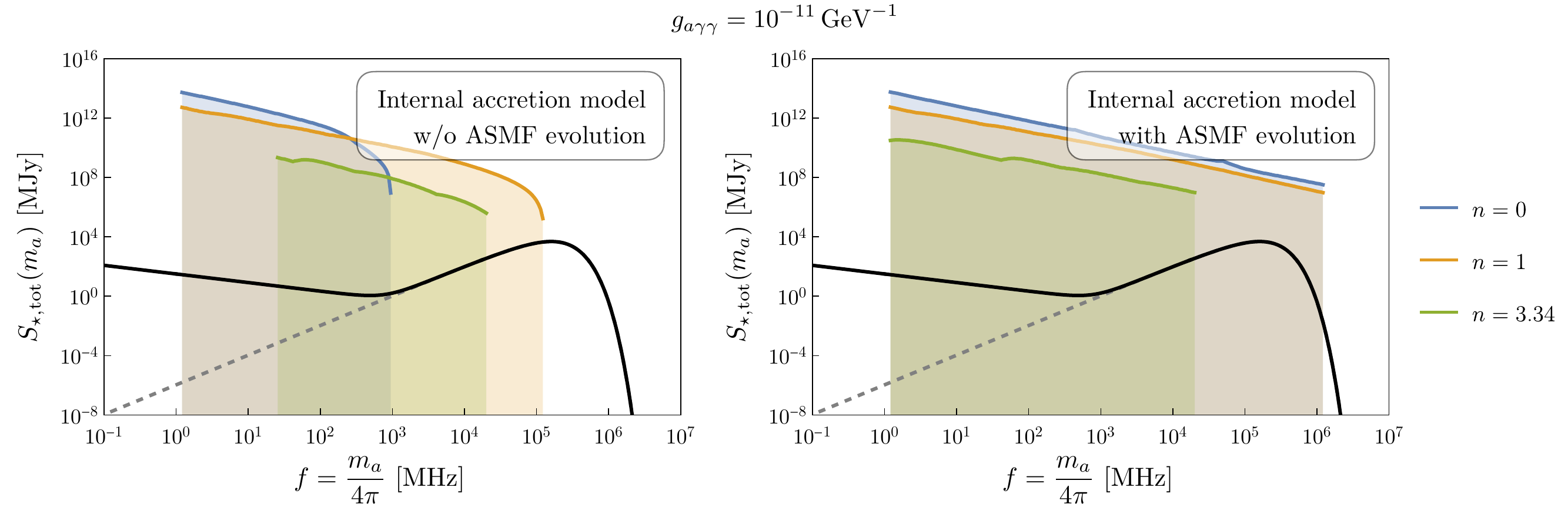}
    \end{subfigure}
    \caption{Spectral flux density $S_{\star,{\rm tot}}(m_a)$ for $g_{a\gamma\gamma}=10^{-11}\,{\rm GeV}^{-1}$ in the unit of MJy with a central frequency corresponding to each axion mass $f=m_a/(4\pi)$.
    Coloured lines represent different temperature dependence of the axion potential, $m_a(\,T>(m_{a,0}f_a)^{1/2})=m_{a,0}[(m_{a,0}f_a)^{1/2}/T]^n$, for which $n=0$ (blue), $n=1$ (orange), and $n=3.34$ (green).
    As mentioned below Eq.~\eqref{eq:m_T}, $m_a$ without a temperature argument denotes $m_{a,0}$, i.e., $m_a=m_{a,0}$.
    The black-solid line describes the radio backgrounds, including the cosmic microwave background (dashed) and observed radio excesses, parameterized as $T_{\rm bkg} = T_{\rm CMB} + 30.4\,{\rm K}\left(f/310\,{\rm MHz}\right)^{-2.58}$ \cite{Dowell:2018mdb}.
    Each panel illustrates a different accretion model, as detailed in Sec.~\ref{sec:accmodels}.}
    \label{fig:SigPlot}
\end{figure}

Let us introduce the spectral flux densities $S_{\star,{\rm tot}}$, derived from the total flux densities $F_{\star,{\rm tot}}$ discussed in Sec.~\ref{sec:accmodels}.
$F_{\star,{\rm tot}}$ represents the power integrated over solid angles per unit area at detection, without containing spectral information.
Since the emitted photons from the parametric-resonant ASs are nearly monochromatic at the central frequency $f\simeq m_a/(4\pi)$, we can construct the spectral flux density $S_{\star,{\rm tot}}$ by dividing $F_{\star,{\rm tot}}$ by the frequency width.
The dominant effect broadening the monochromatic signal is the Doppler shift caused by the relative velocities between the earth and the AS.
We account for this relative velocity as a dark matter velocity dispersion, $v_{\rm rel}\simeq v_{\rm vir}\simeq \mathcal{O}(200 - 300)\,{\rm km/s}$, which results in $\Delta f/f \simeq v / c \simeq 10^{-3}$.
The resulting spectral flux density $S_{\star,{\rm tot}}$ is thus given by
\begin{align}
    S_{\star,{\rm tot}}\left(m_a=4\pi f\right)\simeq\frac{F_{\star,{\rm tot}}(m_a)}{10^{-3}\, m_a/(4\pi)}
    .
    \label{eq:Stot}
\end{align}

We calculate the spectral flux density $S_{\star,{\rm tot}}$ as a function of the axion mass $m_a$ for each accretion model in Fig.~\ref{fig:SigPlot}.
We scan the axion mass from $m_a=10^{-8}\,{\rm eV}$ to $m_a=10^{-2}\,{\rm eV}$, motivated by the radio window for terrestrial radio telescopes rather than by our signals from AS accretion.
The radio window ranges from 10\,{MHz}, limited by the ionosphere, to around 10 - 100\,GHz, constrained by atmospheric absorption.
We scan the axion mass to encompass this range.
Since this radio window applies to terrestrial observations, we could explore frequencies below 10\,MHz using space-borne telescopes.
As a benchmark, we choose $g_{a\gamma\gamma} = 10^{-11}\,{\rm GeV}^{-1}$.
Fig. \ref{fig:SigPlot} shows the height of the spectral signals for a given axion mass in MHz.
Each coloured line shows the different temperature dependence of the axion potential as given in Eq.~\eqref{eq:m_T}: $m_a(T)=m_{a,0}[(m_{a,0}f_a)^{1/2}/T]^n$ for $T>(m_{a,0}f_a)^{1/2}$, and $m_a(T)=m_{a,0}$ for $T\leq (m_{a,0}f_a)^{1/2}$.
We consider $n=0$ (blue), $n=1$ (orange), and the QCD-axion-like $n=3.34$ (green).
The black-solid line represents the combined radio backgrounds of the cosmic microwave background and observed radio excesses, parameterized by a simple fitting formula for the temperature: $T_{\rm bkg} = T_{\rm CMB} + 30.4\,{\rm K}\left(f/310\,{\rm MHz}\right)^{-2.58}$ \cite{Dowell:2018mdb}.
The black-dashed line indicates the cosmic microwave background component of the total radio backgrounds.

We observe that the signal strength depends mainly on the accretion model and the axion parameters $m_a$ and $n$.
In general, the internal accretion scenario produces larger signals than the external accretion scenario.
As expected, the maximal external accretion model yields larger signals compared to the external accretion model, attributed to the absence of the suppression factor $(\mathcal{M}/\mathcal{M}_{h,{\rm min}})^{-2/3}$.
The signals are dominated by the high-mass tail of the ASMF, given that $S_{\star,{\rm tot}} \propto \int d M_\star (dn_\star/dM_\star)(\delta M_\star/\delta t) \simeq \int d\mathcal{M} (dn/d\mathcal{M})(\delta M_\star/\delta t) \propto \int d\mathcal{M} \mathcal{M}^{-3/2}\mathcal{M}^{p}$, where $p \simeq 2/3$ (external accretion), $p\simeq 4/3$ (maximal external accretion), and $p\simeq 1$ (internal accretion).
In addition, the ASMF evolution, i.e., the long-time evolution of ASs, boosts the signal in the high axion mass regime and also shifts the axion mass cutoffs for the internal accretion scenario.
We find that the signal tends to decrease as the axion mass increases across the accretion models, which is mainly due to the signal width, $\Delta f \simeq 10^{-3}\,m_a/(4\pi)$, in Eq.~\ref{eq:Stot}.
On top of that, the mass dependence of $F_{\star,{\rm tot}}(m_a)$ adds a small additional dependence through the total number of parametric-resonant ASs, $N_\star^{\rm res}$, and the accretion rates.
The mass cutoffs depend on the axion mass temperature evolution $n$.
Interestingly, due to the large number of resonant ASs, $N_\star^{\rm res}\gtrsim 10^{10}$, across various accretion models and over a broad region of parameter space, and due to the narrow width of their emission, AS accretion produces overwhelming signals compared to the radio backgrounds.

\subsection{Constraints on axion parameters from radio observations\label{sec:flux-constraints}}

\begin{figure}[t]
    \centering
    \begin{subfigure}{\textwidth}
    \includegraphics[width=\linewidth]{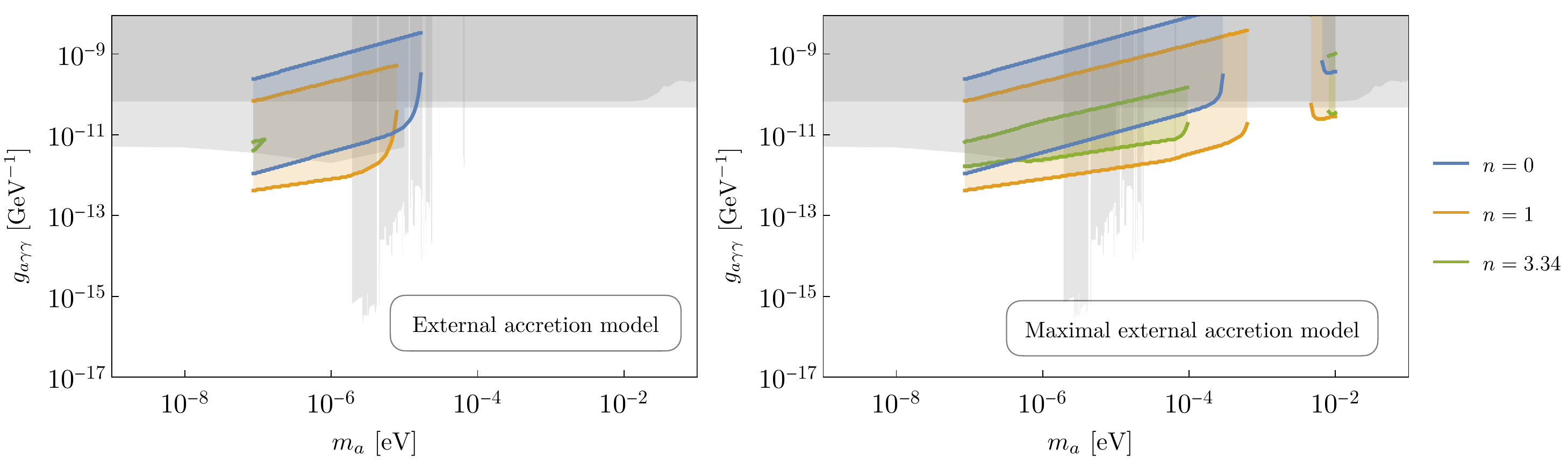}
    \end{subfigure}
    \begin{subfigure}{\textwidth}
    \includegraphics[width=\linewidth]{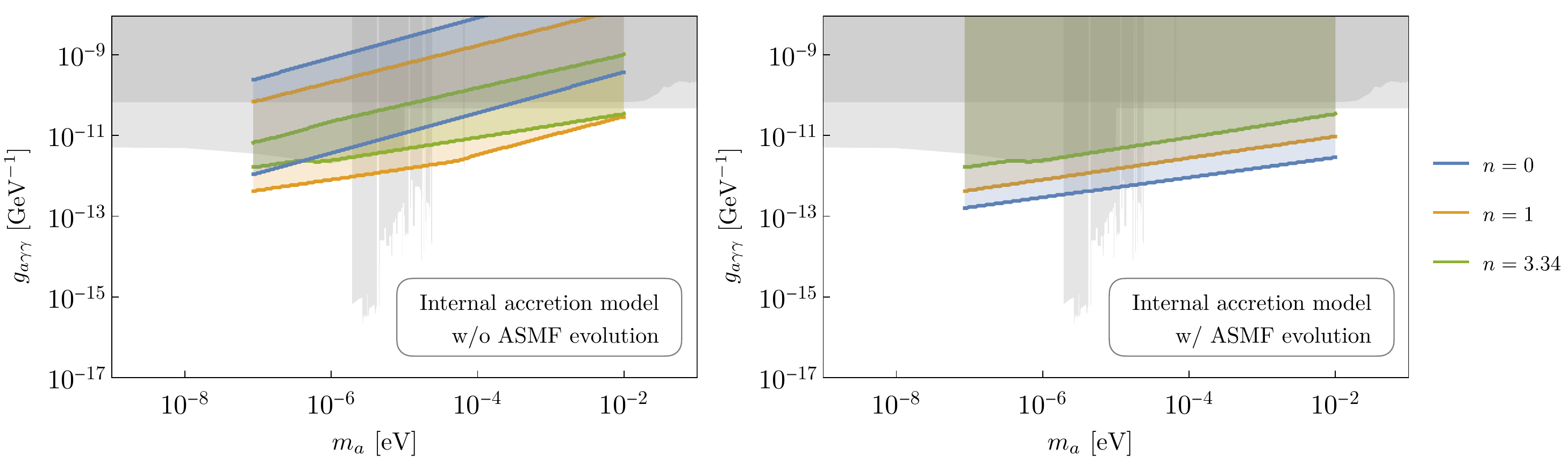}
    \end{subfigure}
    \caption{Exclusion plots for simple background comparison depending on different accretion models.
    The spectral flux density $S_{\star,{\rm tot}}(m_a)$ is compared to the background spectral flux density parametrized by $T_{\rm bkg} = T_{\rm CMB} + 30.4\,{\rm K}\left(f/310\,{\rm MHz}\right)^{-2.58}$.
    The left mass cutoff corresponds to $f=10\,{\rm MHz}$, which is the lower edge of the radio window for terrestrial observations. We scan the axion mass over the range $m_a = 10^{-8}\,{\rm eV} - 10^{-2}\,{\rm eV}$.
    The upper boundary occurs where the decay mass of the AS becomes smaller than the ASMF low-mass cutoff, except for the internal accretion model with ASMF evolution due to the long-time accretion of ASs, in which we allow deviations from the core-halo relation for the initial ASMF.
    Coloured lines represent the temperature dependence of the axion potential, parameterized by $n=0$ (blue), 1 (orange), and 3.34 (green).
    The dark shades correspond to existing constraints from astrophysics, helioscopes, and haloscopes.}
    \label{fig:bkgcomp}
\end{figure}

\begin{figure}[t]
    \centering
    \begin{subfigure}{\textwidth}
    \includegraphics[width=\linewidth]{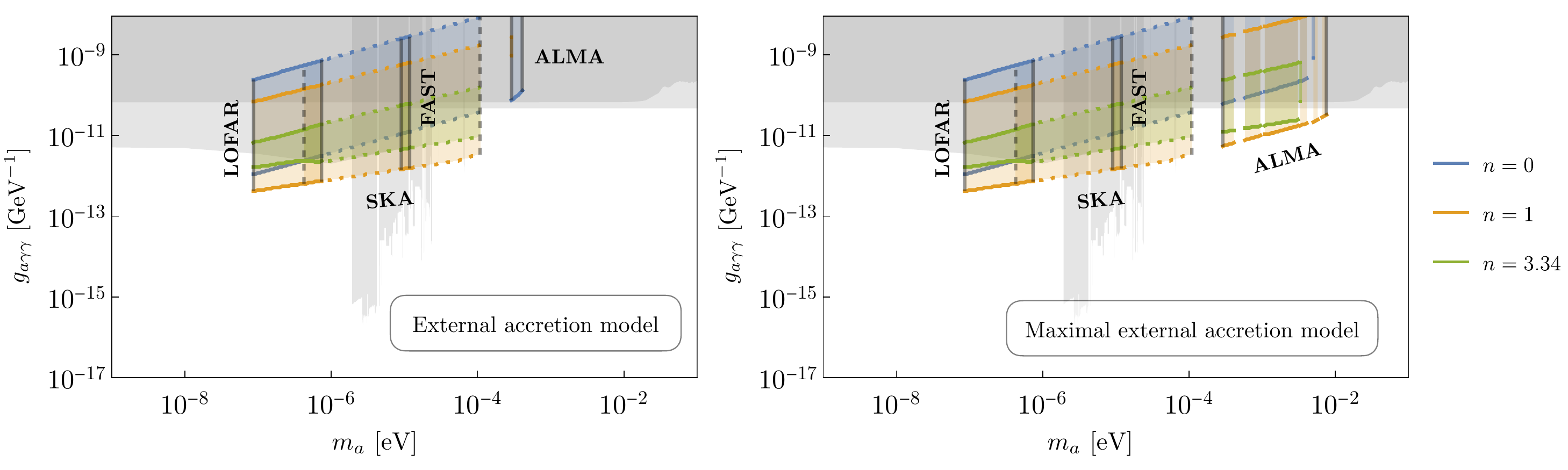}
    \end{subfigure}
    \begin{subfigure}{\textwidth}
    \includegraphics[width=\linewidth]{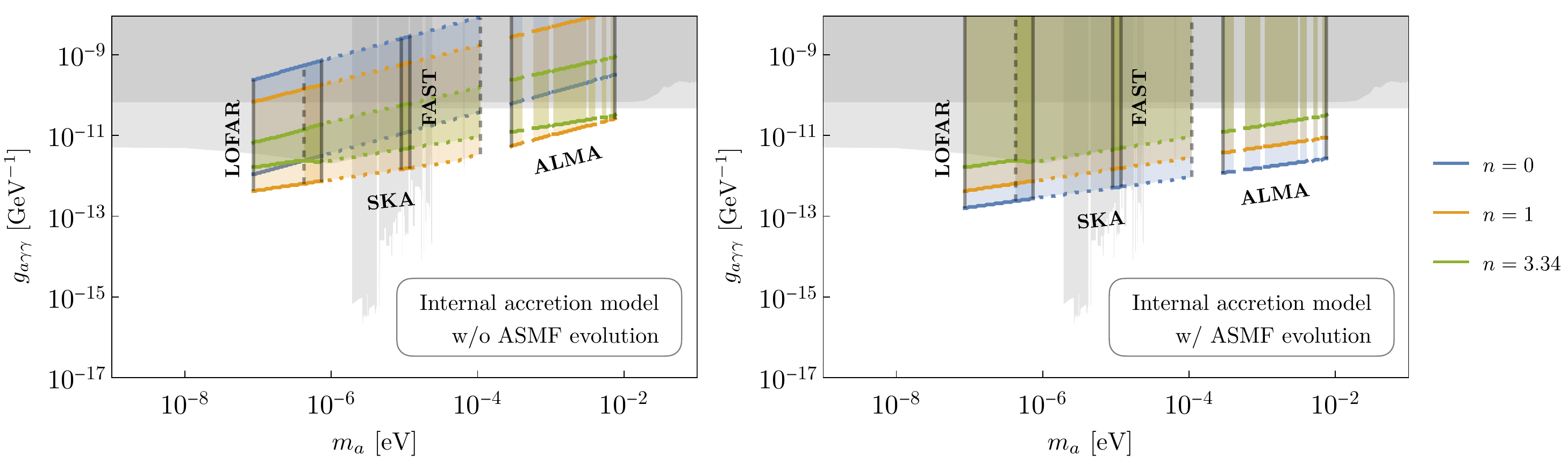}
    \end{subfigure}
    \caption{Exclusion plots comparing our signals with the sensitivities of telescopes---LOFAR, FAST, SKA, and ALMA---across different accretion models.
    The spectral flux density $S_{\star,{\rm tot}}(m_a)$ is translated into the antenna temperature $T_{\rm ant}^{\rm pb}$ and compared to the minimum observable temperature $T_{\min}$.
    SKA constraints are denoted by dotted lines, while all others are represented by solid lines.
    ALMA constraints appear as narrow strips due to the six sets of frequency bands, which have gaps between them.
    The low- and high-mass cutoffs are mostly related to the frequency coverage of the telescopes and sensitivities as well.
    A space-borne radio telescope can search for axion masses below $10^{-7}\,{\rm eV}$.
    The upper boundary excludes the regime where the AS decay mass is smaller than the ASMF low-mass cutoff in the first three models, whereas the internal accretion model with ASMF evolution includes this regime by allowing deviations from the core-halo relation in the initial ASMF.
    Coloured lines show the temperature dependence of the axion potential, with $n=0$ (blue), 1 (orange), and 3.34 (green).
    The dark shades represent existing constraints from astrophysics, helioscopes, and haloscopes.}
    \label{fig:telescopes}
\end{figure}

We perform different approaches to derive constraints on axion parameters, $m_a$ and $f_a$, in this section.
The spectral flux density $S_{\star,{\rm tot}}(m_a)$ in Eq.~\eqref{eq:Stot} is closely connected to the observable in radio telescopes, namely antenna temperature $T_{\rm ant}$ as a representative quantity.
We analyse these signals by directly comparing them to the combined radio backgrounds in Fig.~\ref{fig:bkgcomp}, and by using the sensitivities of radio telescopes in Fig.~\ref{fig:telescopes}.
For direct comparison, if the signal strength exceeds the radio backgrounds for a given mass---i.e., if the coloured lines are above the black solid curve---the corresponding axion masses would be constrained or potentially detectable.
This approach is simplified but yields conservative results that do not depend on the telescope specifications and provides a precursor to the projections.
When using radio telescope sensitivities, we focus on single-dish modes instead of interferometric modes because our signals are diffuse and approximately isotropic.

\textbf{Simple background comparison.} 
We compare our signals $S_{\star,{\rm tot}}(m_a)$ from each accretion model with radio backgrounds parametrized in terms of $T_{\rm bkg} = T_{\rm CMB} + 30.4\,{\rm K}\left(f/310\,{\rm MHz}\right)^{-2.58}$ \cite{Dowell:2018mdb}, and derive the corresponding constraints in Fig.~\ref{fig:bkgcomp}.
The radio background temperature is expressed as spectral flux density using the black-body radiation formula, resulting in the black-solid line in Fig.~\ref{fig:SigPlot}.
As mentioned, if the spectral flux densities from signals, whose peaks are denoted by coloured lines in Fig.~\ref{fig:SigPlot}, exceed those of the radio backgrounds, the excesses could be easily observed.
Based on this criterion, we can derive conservative constraints or identify the detectable parameter region in our accretion models, independent of the specific details of the telescopes.

In Fig.~\ref{fig:bkgcomp}, each accretion model is presented in separate panels.
We scan the axion mass $m_a\in [10^{-8}, 10^{-2}]\,{\rm eV}$ to cover the radio window. 
We impose an additional low-mass cutoff based on the radio window boundary at $f = 10\,{\rm MHz}$, which corresponds to $m_a\simeq 10^{-7}\,{\rm eV}$.
The constraints are determined by complex combinations of factors affecting the signals, as described by Eqs.~\eqref{eq:F_totalext}, \eqref{eq:F_total_high}, \eqref{eq:F_tot_new}, and \eqref{eq:Ftotintlong}.
However, the lower and upper boundaries of the constraints can be easily understood by comparing the mass cutoffs, rather than the signal strengths, which are almost always sufficiently large.

The lower boundaries, from top to bottom, are in the order of $n=0$, $n=3.34$, and $n=1$, except for the internal accretion model with long-time AS evolution.
The lower boundary appears when the AS decay mass in Eq.~\eqref{eq:M_Decay_Gaussian}, which is approximately independent of $n$, becomes too heavy and exceeds the high-mass cutoff of the ASMF.
From Eq.~\eqref{eq:M_h_max}, the high-mass cutoff $\mathcal{M}_{0,\max}(m_a,n)$ is determined by $\mathcal{M}_0(m_a,n)$, which increases monotonically with larger $n$, with ratios on the order of $\sim 10^3$.
Specifically, $\mathcal{M}_0(m_a, n=3.34)/\mathcal{M}_0(m_a,n=1) \simeq \mathcal{M}_0(m_a, n=1)/\mathcal{M}_0(m_a,n=0) \simeq \mathcal{O}(10^3)$ \cite{fairbairn_structure_2018,maseizik_pheno_2024}.
On the other hand, the high-mass cutoff $\mathcal{M}(M_{\star,\max})$, corresponding to the maximum stable AS mass $M_{\star,\max}$ in Eq.~\eqref{eq:M_starmax}, exhibits the opposite trend with respect to $n$, i.e., it decreases with larger $n$ due to smaller $f_a$ for a given $m_a$ \cite{fairbairn_structure_2018,maseizik_pheno_2024}.
$M_{\star,\max}$ truncates the high-mass tail of the ASMF for $n=3.34$, resulting in the final high-mass cutoffs of the ASMF in the order $n=0$, $n=3.34$, and $n=1$ (see the example plot in Fig.~\ref{fig:exampleMF} of App.~\ref{app:ASMF}).
This cutoff order is reflected in the lower boundary order, as shown in Fig.~\ref{fig:bkgcomp} for different accretion models, except for the internal accretion with long-time AS evolution.
In the long-time evolution case, the shift of $\mathcal{M}_\gamma$ to $\mathcal{M}_\gamma \simeq 2 M_{\star,\gamma}$ alters the boundary by affecting the regime where $M_\star(\mathcal{M}_{0,\max})<M_{\star,\gamma} < M_{\star,\max}$. Here, $M_\star(\mathcal{M}_{0,\max})$ denotes the AS mass corresponding to the MC mass cutoff $\mathcal{M}_{0,\max}$ according to the core-halo relation.
In this case, replacing $\mathcal{M}_\gamma = \mathcal{M}(M_{\star,\gamma})$ with $\mathcal{M}_\gamma \simeq 2 M_{\star,\gamma}$ makes ASs resonant in this parameter region, which was not possible previously because the ASs were too light.
In other words, the long-time evolution of ASs increases their mass to be comparable to the host MC masses, allowing them to reach the AS decay mass.
Consequently, the lower boundary appears where $M_{\star,\gamma} \simeq M_{\star,\max}$, in the order of $n=3.34$, $n=1$, and $n=0$.

Next, we discuss the upper boundaries for large $g_{a\gamma\gamma}$ in figure \ref{fig:bkgcomp}.
If the axion-photon coupling is too large, the decay mass of AS becomes too light, according to Eq.~\eqref{eq:M_Decay_Gaussian}, even lighter than any AS of the ASMF, i.e., $\mathcal{M}_\gamma<\mathcal{M}_{\min}(\Phi)$ in the MCMF.
As discussed in detail in Secs.~\ref{sec:CoreHalo_ASMF} and \ref{sec:self-similar-model}, we exclude this parameter regime in the external accretion models and the internal accretion model without ASMF evolution because of adherence to the core-halo relation for the initial ASMF, which results in the upper boundaries.
On the other hand, we relax this restriction and allow configurations straying from the core-halo relation in the internal accretion model with ASMF evolution.
Since the internal accretion model with ASMF evolution lowers $\mathcal{M}_\gamma$ according to Eq.~\eqref{eq:MMCgammalong}, the signals are boosted, and the upper boundary is shifted upwards as a result of relaxing the restriction.
In this regime, a detailed study is required further to verify whether all AS-MC systems are in parametric resonance, taking into account their evolution under the resonance and the backreaction on dark matter abundance throughout cosmological history.
We leave this for future study, although this upper region is already excluded by other constraints, which originate from astrophysical observations, helioscopes, and haloscopes.
We represent these existing bounds in grey shades.

\begin{table}[]
\centering
\resizebox{\textwidth}{!}{
\begin{tabular}{l | c | c | c | c}
    \hline &&&\\[-2.3ex]
    Specifications & \makecell{LOFAR-LBA\\\cite{LOFAR:sens,LOFAR:2013jil}} & \makecell{FAST\\\cite{2020FAST19beam,2020InnovFAST}} & \makecell{SKA-Low /\\ SKA1-Mid\\\cite{SKA:sens,SKA:design,Braun:2019gdo}} & \makecell{ALMA\\\cite{ALMA:sens,ALMA:tech}} \\
    &&&\\[-2.7ex]
    \hline\hline
    &&&\\[-2.3ex]
    Frequency range [MHz] & [10, 90] & [1050, 1450] & \makecell{[50, 350] /\\ \newline [350, 14k]}  &  \makecell{[35k, 50k] (B1),\\ \newline [67k, 116k] (B2 - B3),\\ \newline [125k, 373k] (B4 - B7),\\ \newline [385k, 500k] (B8),\\ \newline [602k, 720k] (B9),\\ \newline [787k, 950k] (B10)} \\
    &&&\\[-2.7ex]
    \hline
    &&&\\[-2.3ex]
    Diameter $D_{\rm tel}$ [m] & 81.34 & 300 & 35 / 15 & 12 \\
    &&&\\[-2.7ex]
    \hline
    &&&\\[-2.3ex]
    Number of telescopes $n_{\rm tel}$ & 40 & 19 & \makecell{911 / 200} & 50 \\
    &&&\\[-2.7ex]
    \hline
    &&&\\[-2.3ex]
    Number of polarizations $n_{\rm pol}$ & 2 & 2 & 2 & 2 \\
    &&&\\[-2.7ex]
    \hline
    &&&\\[-2.3ex]
    Observation time [hrs] & 100 & 100 & 100 & 100 \\
    &&&\\[-2.7ex]
    \hline
    &&&\\[-2.3ex]
    Receiver noise $T_{\rm rcv}$ [K] & $10^4$ & 24 & \makecell{40 / 20} &   \\
    &&&\\[-2.7ex]
    \hline
    &&&\\[-2.3ex]
    System temperature $T_{\rm sys}$ [K] & $T_{\rm rcv} + T_{\rm bkg}$ &  $T_{\rm rcv} + T_{\rm bkg}$ &  $T_{\rm rcv} + T_{\rm bkg}$ &   \makecell{90 (B1),\\200 (B2 - B3),\\400 (B4 - B7),\\870 (B8),\\2200 (B9),\\3700 (B10)} \\
    &&&\\[-2.7ex]
    \hline
    &&&\\[-2.3ex]
    Effective area $A_{\rm eff}$ & $\min(16\lambda^2,~4500\,{\rm m}^2)$ & $0.55\, \pi(D_{\rm tel}/2)^2$ & $0.8\, \pi(D_{\rm tel}/2)^2$ & $0.26\, \pi(D_{\rm tel}/2)^2$ \\
    &&&\\[-2.7ex]
    \hline
    &&&\\[-2.3ex]
    Primary beam angle $\theta_{\rm pb}$ & $\pi/3$ & $1.02 (\lambda/D_{\rm tel})$ & $1.02 (\lambda/D_{\rm tel})$ & $1.02 (\lambda/D_{\rm tel})$ \rule[-0.9ex]{0pt}{0pt}\\
    \hline
\end{tabular}}
\caption{Specifications of three existing telescopes and one projected telescope, SKA, used in our analysis are provided.
The chosen parameters may not maximize detectability of our signals but still offer comparable estimates of sensitivity.
LOFAR-LBA refers to an array of low-band antennas (LBA), not a dish-type telescope.
The number of telescopes for LOFAR-LBA refers to the number of core and remote stations.
We extract the receiver noise from data on the ratio between background temperature and system temperature for the LOFAR LBA system.
Due to the low frequency range of LOFAR-LBA, the LBA dipole beam covers almost the entire sky above $30^\circ$ elevation; we conservatively set its value to $\pi/3$.
FAST is a single-dish telescope, where $n_{\rm tel}$ represents the number of beams.
SKA is a projected array of telescopes, and we consider SKA-Low and SKA1-Mid configurations.
For ALMA, we benchmark the parameters provided in the sensitivity calculator.
B1 - B10 denote 10 different bands in the ALMA system.
The system temperature for ALMA is directly taken from the sensitivity calculator in an attempt to derive conservative values.
The effective areas of all the telescopes indicate the efficiency factors relative to their geometric areas; for example, 0.26 for ALMA includes quantization efficiency, correlator efficiency, and aperture efficiency.
The primary beam angle is chosen to be the half-power beam width. 
See \cite{Ferreira:2024ktd,Guo:2024oqo,Caputo:2018vmy,Arza:2024iuv} for recent applications of these telescopes in phenomenological studies.}
\label{tab:telescopes-specs}
\end{table}

\textbf{Radio telescope sensitivities.}
We perform a similar analysis using radio telescopes to detect AS radio lines, again employing single-dish modes.
Since our signals are diffuse and approximately isotropic, single-dish modes are preferred over interferometric modes.
Interferometric modes utilize an array of dishes, where any combination of two dishes acts as a single dish with a radius equivalent to the baseline, i.e., the distance between the two dishes.
These interferometric modes are designed for better angular resolution and eliminate angular scales larger than those of the shortest baseline, which is a downside for our signals.
Therefore, we consider only single-dish modes in the following analysis.

In radio astronomy, observables are often defined in terms of temperatures.
Let us start by considering a single dish.
Our intrinsic signal, the spectral flux density $S_{\star,{\rm tot}}$, can be converted into the spectral flux density observed by the telescope, $S_{\star,{\rm pb}}(m_a)$, by factoring in two experimental parameters: the bandwidth $\Delta B$ and the field of view of the primary beam $\Delta \Omega_{\rm pb}$.
$S_{\star,{\rm pb}}(m_a)$ is, in turn, translated into an antenna temperature for the primary beam, $T_{\rm ant}^{\rm pb}$.
First, if the telescope has a frequency bandwidth larger than the intrinsic frequency dispersion of the signal $\Delta f$, the observed signal $S_{\star,{\rm pb}}(m_a)$ should be suppressed by a fraction of $\Delta f/\Delta B \simeq 10^{-3}\,m_a/(4\pi\Delta B)$.
Since current radio telescopes are capable of more than 1000 channels, throughout this work, we set $\Delta B \simeq \Delta f$.
Secondly, when the single-dish is pointed at the sky, it covers only a small portion of the solid angle.
We thus multiply $S_{\star,{\rm tot}}(m_a)$ by the fraction of solid angles $\Delta\Omega_{\rm pb}/(4\pi)$ to estimate the spectral flux density within the solid angle coverage of the primary beam, $\Delta\Omega_{\rm pb}$:
\begin{align}
    S_{\star,{\rm pb}}(m_a)
    \simeq 
    S_{\star,{\rm tot}}(m_a) \frac{\Delta\Omega_{\rm pb}}{4\pi},
\end{align}
where $\Delta\Omega_{\rm pb} = 2\pi(1-\cos(\theta_{\rm pb}/2))$.
The primary beam angle $\theta_{\rm pb}$ is given in Table~\ref{tab:telescopes-specs}.
The corresponding antenna temperature for the primary beam is defined as
\begin{align}
    T_{\rm ant}^{\rm pb}
    &=
    \frac{1}{2k_{\rm B}} A_{\rm eff}\,S_{\star,{\rm pb}}(m_a),
\end{align}
where $A_{\rm eff} = \eta A$ is the effective area of the telescope, with $\eta$ being an efficiency factor and $A$ the physical area of the dish.
The effective area for each telescope is provided in Table~\ref{tab:telescopes-specs}.

The noise counterpart of the antenna temperature, referred to as the minimum observable temperature, is defined by
\begin{align}
    T_{\rm min}
    &=
    \frac{T_{\rm sys}}{\displaystyle\sqrt{\Delta B\, t_{\rm obs}}},
\end{align}
where
$T_{\rm sys} = T_{\rm rcv}+T_{\rm bkg}$ is the system temperature as a sum of the receiver noise temperature $T_{\rm rcv}$, provided in Table~\ref{tab:telescopes-specs} for each telescope, and the radio-background temperature $T_{\rm bkg}=T_{\rm CMB}+30.4\,{\rm K}\left(f/310\,{\rm MHz}\right)^{-2.58}$ \cite{Dowell:2018mdb}.
$\Delta B$ is the bandwidth that is set equal to the width of the radio-line signal due to the Doppler shift, i.e., $\Delta B \simeq \Delta f \simeq 10^{-3}\, m_a / (4\pi)$.
$t_{\rm obs}$ is the duration of observation, which we set to 100 hours.
The signal-to-noise ratio for a single-dish telescope is given by
\begin{align}
    \left(\frac{S}{N}\right)_{\rm single}
    &=\frac{T_{\rm ant}^{\rm pb}}{T_{\rm min}}.
\end{align}
Because each telescope usually consists of multiple single dishes, the signal-to-noise ratio for an array of $n_{\rm tel}$ dishes operating in the single-dish mode with $n_{\rm pol}=2$ polarizations is
\begin{align}
    \left(\frac{S}{N}\right)_{\rm array}
    &=\sqrt{n_{\rm tel} \, n_{\rm pol}}
    \left(\frac{S}{N}\right)_{\rm single},
    \\
    &=
    \sqrt{n_{\rm tel} \, n_{\rm pol}}
    \frac{T_{\rm ant}^{\rm pb}}{T_{\rm min}},
    \label{eq:SNRarray}
\end{align}
where the corresponding telescope specifications are summarized in Table~\ref{tab:telescopes-specs}.

The constraint plot based on radio-telescope sensitivities is presented in Fig.~\ref{fig:telescopes}.
We display the region where the signal-to-noise ratio in Eq.~\eqref{eq:SNRarray} is larger than 1, i.e., $(S/N)_{\rm array} > 1$.
Even though the signal $S_{\star,{\rm tot}}$ is smaller than the radio backgrounds, e.g., in the external accretion model for $n=3.34$ in Fig.~\ref{fig:SigPlot}, it can still be detected by radio telescopes due to its narrow spectral shape, which contrasts with the mostly broad and smooth backgrounds and foregrounds.
As in Fig.~\ref{fig:bkgcomp}, each panel represents a different accretion model, with coloured lines having different temperature dependence of the axion potential.
The dark shades denote the existing constraints from astrophysics, helioscopes, and haloscopes.\footnote{Axion haloscopes rely on the local dark matter abundance. If most of the axion dark matter is bound in MCs, then the haloscope constraints could be significantly modified, depending on the encounter rate between the earth and MCs \cite{Hogan:1988mp}. On the other hand, axion streams resulting from the tidal disruption by stellar objects could replenish the local dark matter density around the earth. See \cite{OHare:2023rtm} for a recent discussion. We thank J\'er\'emie Quevillon and Enrico Nardi for pointing this out.}
One notable feature is the practical independence of the constraints from the accretion models.
While the signal strength is significantly influenced by the accretion models, the more rapid dependence on $g_{a\gamma\gamma}$---which affects the sheer existence of parametric-resonant AS-MC systems through ASMF cutoffs---leads to the sharp behaviour of the signals, such as an on-off pattern, depending on $g_{a\gamma\gamma}$.
In other words, the boundaries are primarily determined by the ASMF cutoffs, not by setting $(S/N)_{\rm array}$ equal to 1.
As a result, the constraint boundaries become similar.
Even the most conservative accretion model, i.e., the external accretion model, generates sufficiently strong signals to be detected by radio telescopes.
Without a doubt, the signal-to-noise ratios differ across the accretion models within the similar constrained regions, as the signal strength highly depends on the accretion models.
The upper boundary appears where $g_{a\gamma\gamma}$ is so large that the corresponding AS decay mass becomes lighter than any AS of the ASMF, i.e., $\mathcal{M}_\gamma<\mathcal{M}_{\min}(\Phi)$ in the MCMF, as discussed in the simple background comparison.
However, in the case of the internal accretion model with ASMF evolution due to long-time AS accretion, relaxing the core-halo relation for the initial ASMF shifts the upper boundary upwards, and lowering $\mathcal{M}_\gamma$ according to Eq.~\eqref{eq:MMCgammalong} strengthens the signals by making more AS-MC systems resonant.
The low- and high-mass cutoffs are determined from frequency ranges of telescopes.

Although detecting signals more effectively may require detailed strategies, such as 1) using spatial anisotropy to identify differences, 2) pointing radio telescopes towards galactic poles, and 3) performing careful analysis to reduce foregrounds and backgrounds around atomic- and molecular-transition frequencies including 21\,cm,
our models provide an alternative way to search for axion signals around $m_a \simeq 10^{-7} - 10^{-2}\,{\rm eV}$ and potentially even a broader range.
Accordingly, our models also motivate the space-borne detection of radio signals, and spectral-line searches to detect possible axion radio signals.


\section{Discussion and outlook\label{sec:discussion}}

We have demonstrated different accretion mechanisms for solitonic ASs and explored its phenomenological consequences through parametric resonance induced by the axion-photon coupling.
In the post-inflationary scenario, axion dark matter forms substructures such as MCs and ASs over cosmological evolution, with ASs expected to be hosted by MCs.
At the same time, if the AS density is sufficiently large, parametric resonance via axion-photon coupling could be induced.
We have estimated the number of parametric-resonant AS-MC systems in our galaxy, using the parametrization of the MCMF based on analytic Press-Schechter formalism and the core-halo relation.
Based on simulation results, we assume that 75\% of the dark matter is bound in MCs, while our results should not be very sensitive to the bound fraction.

Given the distribution of AS-MC systems in our galaxy, we have examined four different accretion models, categorized into two scenarios: external accretion and internal accretion.
In external accretion models, the host MC first captures background dark matter, which is then transferred to the AS core through virialization of the AS-MC system.
For internal accretion, the (isolated) host MC directly sources the mass growth of its AS core, whose evolution can be described by a simple fitting formula based on numerical simulations and analytical modeling.
In each accretion model, we have calculated the mass growth rate and converted it to the spectral flux density $S_{\star,{\rm tot}}$ emitted from the resonant ASs, depending on axion model parameters such as axion mass $m_a$, axion-photon coupling $g_{a\gamma\gamma}$, and the temperature dependence of the axion potential, parameterized by $n$.
We have examined the detectability of these signals through comparisons with radio backgrounds and radio telescope sensitivities.
Considering the radio window for terrestrial telescopes, we focused on the axion mass range $m_a = 10^{-8} - 10^{-2} \,{\rm eV}$.

The signals from resonant ASs in our galaxy are diffuse and approximately isotropic, making single-dish modes of radio telescopes more suitable for detection than interferometric modes.
In addition, the photons emitted from decaying axions due to parametric resonance are nearly monochromatic up to the Doppler shift caused by dark matter velocity dispersion, because axions within ASs are non-relativistic.
The narrow line signals are expected to help reduce foregrounds and backgrounds.
Lastly, and most importantly, the dependence of the signals on $g_{a\gamma\gamma}$ is dominated by ASMF cutoffs, resulting in an on-off pattern as $g_{a\gamma\gamma}$ varies.
This feature is reflected in the constraint plots, which show a similar shape across different accretion models.

We have derived constraint plots in the $m_a$ and $g_{a\gamma\gamma}$ plane for each accretion model by comparing the signals with radio backgrounds and the sensitivities of three existing telescopes---LOFAR, FAST, and ALMA---and one projected telescope, SKA.
While the signal-to-noise ratio does vary significantly with different accretion models, the resulting constraints exhibit similar shapes because of the ASMF cutoff dependence, as discussed above.
In other words, the independence of constraints from specific accretion models consolidates the coverage of detectable parameters resulting from the accretion of resonant ASs in our galaxy.
The relevant parameter range is $m_a \simeq 10^{-7} - 10^{-2}\,{\rm eV}$ and $g_{a\gamma\gamma} \simeq 10^{-13} - 10^{-9}\,{\rm GeV}^{-1}$.

AS accretion offers a complementary way to observe neV to meV axions in the post-inflationary scenario, where axion dark matter forms substructures, in place of haloscope constraints.
The haloscope constraints are weakened in general as the bound fraction of axion dark matter increases.
Furthermore, our study encourages spectral line searches for axions using radio telescopes.
The diffuse feature makes foreground and background reduction even easier by pointing radio telescopes away from stellar sources, for example, towards the galactic poles, or by utilizing anisotropy from the halo profile.
Finally, space-borne radio telescopes could provide an interesting opportunity to extend the mass range beyond the radio window, particularly to lower masses, $m_a\lesssim 10^{-7}\,{\rm eV}$.

On the theory side, our simplified accretion models already capture the dominant features of AS accretion.
However, the accretion modeling could be improved by incorporating parameter dependence more carefully, such as $\Phi$ in the internal accretion models, and by refining the evolution of the initial ASMF through cosmological history with particular attention to the time evolution of the core mass.
Moreover, developing a hybrid accretion model that simultaneously considers both external and internal accretion would be valuable for future investigation.
It is also worthwhile to note that the population of ASs and MCs would have a distribution with varying AS and MC profiles depending on the parameters.
For the resonance dynamics, exploring parametric-resonant AS evolution within MCs and radio conversion efficiency in detail would provide deeper insights into signals.
Finally, examining our accretion models in pre-inflationary scenarios involving different axion dark matter production mechanisms, such as large misalignment or kinetic misalignment with fragmentation, presents another promising direction for further study.\footnote{This possibility was suggested by G\'{e}raldine Servant in private communications.}

\section*{Acknowledgments}
We would like to thank Dieter Horns, Thomas Schwetz, Samuel J. Witte, Virgile Dandoy, J\'er\'emie Quevillon, Enrico Nardi, Francesca Calore, Marcus Br{\"u}ggen, and G\'{e}raldine Servant for useful discussions and comments.
HS appreciates the hospitality of LAPTh during the visit.
This work is funded by the Deutsche Forschungsgemeinschaft (DFG, German Research Foundation) under Germany’s Excellence Strategy -- EXC 2121 ``Quantum Universe'' -- 390833306.
This article is based upon work from the COST Action COSMIC WISPers CA21106, supported by COST (European Cooperation in Science and Technology).

\appendix
\section{Axion dark matter abundance in the post-inflationary scenario\label{app:DM_Basics}}

In the post-inflationary scenario, where the axion field obtains its value after inflation, the axion fields within a single Hubble volume exhibit a spatial distribution of patches with random field values. 
The axion field begins to evolve around the oscillation temperature $T_{\rm osc}$, defined by $3H(T_{\rm osc}) \simeq m_a(T_{\rm osc})$.
The Hubble parameter is determined by the first Friedmann equation,
\begin{align}
3 H(T)^2 M_{\rm Pl}^2 &= \frac{\pi^2}{30} g_{\star,{\rm R}}(T) T^4, \label{eq:Friedmann}
\end{align}
where $M_{\rm Pl}$ is the reduced Planck mass, and the relativistic degrees of freedom $g_{\star,{\rm R}}(T)$ are obtained from the fit in \cite{Wantz:2009it}.
The temperature dependence of axion mass is given in Eq.~\eqref{eq:m_T}.
For each model parameter ($m_{a,0}, f_a, n$), we can find the oscillation temperature $T_\mathrm{osc}$.
In addition, we require the model parameters to explain the relic abundance of dark matter.
The relic abundance of axion dark matter in the post-inflationary scenario is summarized in a simple equation \cite{fairbairn_structure_2018}: 
\begin{align}
\Omega_a(f_a) &= \frac{1}{6 H_0^2 M_{\rm Pl}^2}\left(1+\beta_{\mathrm{dec}}\right) \frac{c_n \pi^2}{3} m_{a,0} \,m_a(T_{\mathrm{osc}}) f_a^2\left[\frac{a(T_{\mathrm{osc}})}{a_0}\right]^3, \label{eq:relic_abundance}
\end{align}
where $\beta_\mathrm{dec}=2.48$ represents the contribution from decay of the axion string-wall network, calculated from numerical simulations \cite{Kawasaki:2014sqa} (see \cite{Gorghetto:2020qws,Buschmann:2021sdq,Kim:2024wku} for other numerical results), and the subscript 0 denotes the values in the present universe.
The anharmonicity coefficient $c_n$ depends on the temperature index $n$, $c_n = (2.7,2.3,2.1)$ for $n=(0,1,3.34)$ \cite{fairbairn_structure_2018}.
We fix the decay constant $f_a$ to produce the correct relic abundance $\Omega_a h^2=0.12$ for each axion model parameter ($m_{a,0},n$).

An additional constraint on $f_a$ arises from the tensor-to-scalar ratio.
According to observational data constraining the tensor-to-scalar ratio to $r_t <0.07$ \cite{BICEP2:2015xme,fairbairn_structure_2018} (see also \cite{Planck:2018vyg,Planck:2018jri} for recent results),
the Hubble scale of the inflation is constrained to $H_I/(2\pi) = M_{\rm pl} \sqrt{A_s r_t / 8} < 8.2\times 10^{12}\,{\rm GeV}$, where $A_s \sim 10^{-9}$ is the amplitude of the primordial scalar perturbations. 
Since the post-inflationary scenario requires the PQ-breaking scale $f_a$ to be smaller than the inflationary Hubble scale $H_I$, the constraint on the inflationary Hubble scale results in $f_a < 8.2\times 10^{12}\,{\rm GeV}$ \cite{Marsh:2015xka,fairbairn_structure_2018}. 
This condition introduces a cut-off in the low-$m_a$ region, which is considered for any $n$ and $m_a$ in \cite{maseizik_pheno_2024}.

\section{Examples of mass functions and a consistency requirement\label{app:ASMF}}

\begin{figure}[H]
    \centering
    \includegraphics[width=.49\textwidth]{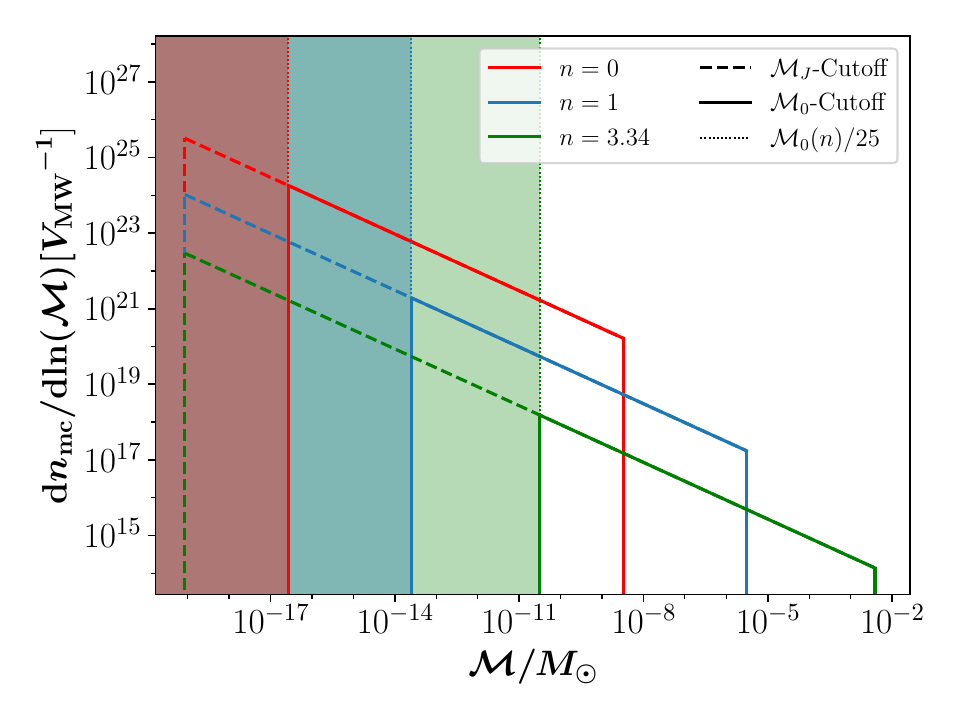}
    \includegraphics[width=.49\textwidth]{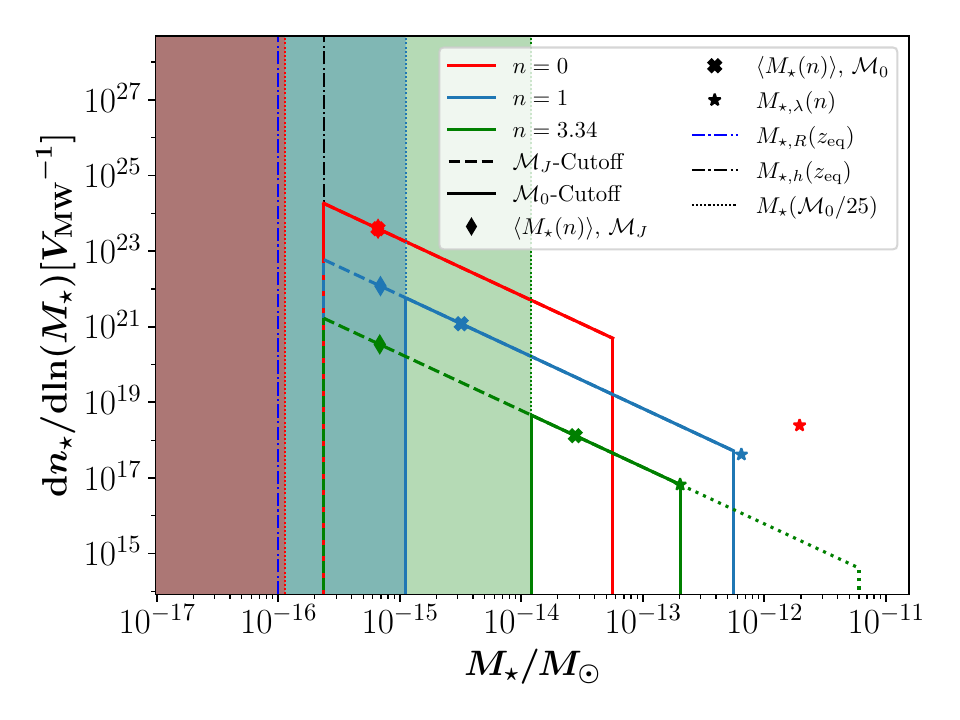}
    \caption{Example plots for the galactic MCMF (left) and ASMF (right) for $m_a = 50\,\mu{\rm eV}$, taken from \cite{maseizik_pheno_2024}.
    Solid lines are the resultant MCMFs and ASMFs after applying various cutoffs represented in dashed, dotted, and dot-dashed lines. In this paper, we discuss the relevant cutoffs in the main text.}
    \label{fig:exampleMF}
\end{figure}

We present example plots for MCMF and ASMF in Fig.~\ref{fig:exampleMF} for $m_a = 50\,\mu{\rm eV}$.
The derivation of these MFs is summarized in Secs.~\ref{sec:MCMF} and \ref{sec:CoreHalo_ASMF}. For more details, refer to \cite{maseizik_pheno_2024}.
For convenience, we also summarize the parameters used in this paper in Table~\ref{tab:Params}.

For completeness, we have incorporated a consistency requirement for our accretion models in numerical calculations, which is a necessary condition for the kinetic regime ensuring the applicability of the self-similar attractor model \cite{levkov_gravitational_2018, Dmitriev:2023ipv}:
\begin{align}
    \mathcal{R} ~ > ~\lambda_{\rm dB} ~ \simeq ~ \frac{2 \pi \hbar}{m_a v_\mathrm{esc}(\mathcal{M})},
\end{align}
which corresponds to $\epsilon\lesssim 1$ in Eq.~\eqref{eq:epsilon}.

\begin{table*}[t]
    \centering
    \resizebox{\textwidth}{!}{
    \begin{tabular}{|c|c|c|}
    \hline
        Quantity & Explanation & Definition \\
        \hline
        $\mathcal{M}$   &   MC mass parameter   & \\
        $\mathcal{R}$   & Spherically homogeneous MC radius & \eqref{eq:R_mc}   \\
        $\mathcal{M}_{0}$   & Characteristic MC mass & \eqref{eq:M0}   \\
        $\mathcal{M}_{0,\min}$   & Applied low-mass MCMF cutoff at $z=0$ & \eqref{eq:M0_cutoff} \\
        $\mathcal{M}_{0,\max}$   & Applied high-mass MCMF cutoff at $z=0$  & \eqref{eq:M_h_max}   \\
        $\mathcal{M}_{h,\min}$   & Corresponding MC mass of ASMF cutoff from core-halo relation & \eqref{eq:M_h_min_CoreHalo} \\
        $\mathcal{M}_{R,\min}$   & Corresponding MC mass of ASMF radius cutoff where $\mathcal{R} = R_\star$ &  \eqref{eq:M_h_min_RadiusCutoff} \\
        $\mathcal{M}_{\min}$   & Low-mass cutoff of MCMF for consistent AS-MC systems   &  \eqref{eq:MClowconsistent}    \\
        $\mathcal{M}_{\gamma,\min}$   & Lower MC mass from combined AS cutoffs for resonant AS-MC systems & \eqref{eq:M_h_min_total} \\
        $\mathcal{M}_{\gamma,\max}$   & Higher MC mass from combined AS cutoffs for resonant AS-MC systems & \eqref{eq:M_h_max_total} \\
        $\mathcal{M}_{\mathrm{tot}}$   & Total mass of MCs in the MW & \eqref{eq:MMCtot} \\
        $\mathcal{M}_{\gamma}$   & MC mass derived from $M_{\star,\gamma}$, depending on accretion models & \eqref{eq:CoreHalo}, \eqref{eq:MMCgammalong} \\
        \hline
        $M_\star$   &   AS mass parameter   &   \\
        $R_\star^{90}$  &   AS radius enclosing 90\% of the AS mass &   \eqref{eq:g_alpha} \\
        $M_{\star,\max}$   & Maximum stable AS mass imposed by self-interactions & \eqref{eq:M_starmax}\\
        $R_{\star,\min}^{90}$   & Minimum stable AS radius imposed by self-interactions & \eqref{eq:R_crit} \\ 
        $M_{\star,\gamma}$   & Decay mass of ASs triggering parametric resonance & \eqref{eq:M_Decay_Gaussian} \\
        $N_\star^{\rm res}$ &   Total number of parametric-resonant ASs in the MW   &   \\
        \hline
    \end{tabular}}
    \caption{Parameters for axion miniclusters (top) and axion stars (bottom) used in this paper.}
    \label{tab:Params}
\end{table*}

\section{Density parameter distribution and survival probability under tidal disruption\label{app:densitypar}}

We consider a broad range of the initial overdensity parameter $\Phi = (\rho_a-\langle\rho_a\rangle) / \langle\rho_a\rangle|_{T_{\rm dec}<T<T_{\rm osc}} \in [10^{-3}, 10^4]$, where $\langle\rho_a\rangle$ is the spatial average of the axion energy density, and $T_{\rm dec}$ is the decoupling temperature of the MC from cosmic expansion, as in \cite{Kolb:1993zz,kolb_large-amplitude_1994,fairbairn_structure_2018}.
The initial cumulative distribution of MCs with $\Phi > \Phi_0$ shows well-fitted by a Pearson-VII-type distribution \cite{kolb_femtolensing_1996,fairbairn_structure_2018},
\begin{align}
\mathcal{P}\left(\Phi>\Phi_0\right)\simeq\frac{1}{\left[1+\left(\Phi_0 / a_1\right)\right]^{a 2}}, \label{eq:CDF}
\end{align}
with $a_1\simeq 1.023$ and $a_2\simeq 0.462$.
From the initial cumulative distribution, we can derive the initial probability distribution: 
\begin{align}
    p_{\Phi,{\rm init}}(\Phi) = \frac{a_2}{a_1 (1 + \Phi/a_1)^{a_2+1}}, \label{eq:PDF}
\end{align}
of MCs as a function of $\Phi$.
For simplicity, we have assumed that the mass distribution of MCs is independent from its density distribution.

\begin{figure}[H]
    \centering
    \includegraphics[width=.49\textwidth]{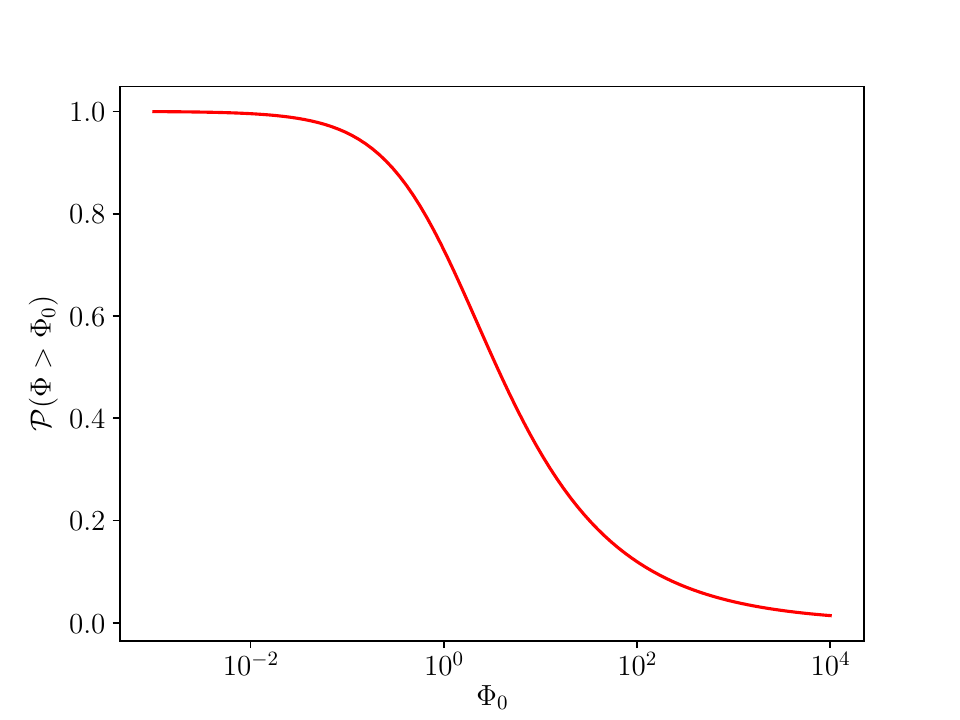}
    \includegraphics[width=.49\textwidth]{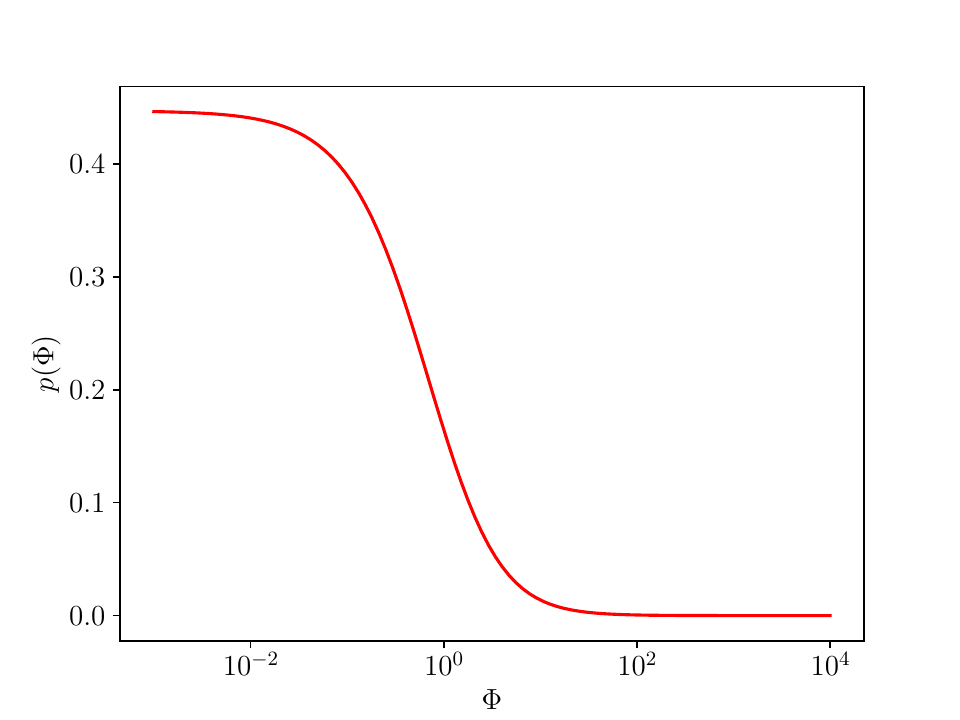}
    \caption{\textit{Left:} Initial cumulative distribution of MCs with $\Phi > \Phi_0$.
    \textit{Right:} Initial probability distribution of MCs with $\Phi$.}
    \label{fig:CDF(Phi)}
\end{figure}

\begin{figure}[H]
    \centering
    \includegraphics[width=.6\textwidth]{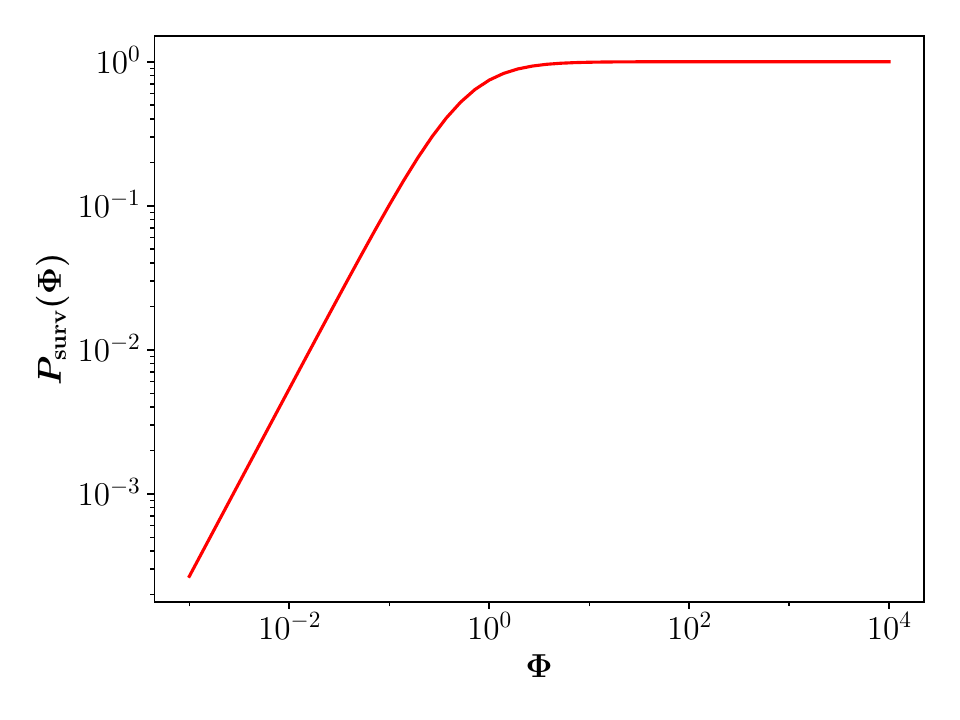}
    \caption{Survival probability of MCs as a function of initial overdensity parameter $\Phi$.}
    \label{fig:P_surv}
\end{figure}

Given the initial distribution of $\Phi$, the tidal disruption further affects the distribution.
The tidal disruption effect can be parametrized into two factors: $R_{\rm surv}$ and $\mathcal{P}_{\rm surv}(\Phi)$.
First, the tidal disruption of MCs in the galactic bulge constrains the galactocentric radius to \cite{kavanagh_stellar_2021}:
\begin{align}
    R ~ & \geq ~R_{\rm surv} \simeq 1\,\mathrm{kpc}.
\end{align}
Secondly, low-density MCs are more easily disrupted, leading to a lower survival probability.
The tidal disruption thus also introduces an additional constraint on the density parameter $\Phi$, with the survival probability of MCs as a function of $\Phi$ \cite{Dandoy:2022prp}:
\begin{align}
    \mathcal{P}_\mathrm{surv}(\Phi) = \frac{1}{2} \left[1 + \tanh\left(\frac{\log_{10}\Tilde{\rho}_\mathrm{MC}(\Phi) - 4.25}{2}\right) \right],
    \qquad
    \Tilde{\rho}_\mathrm{MC}(\Phi) = \frac{\rho_\mathrm{MC}(\Phi)}{M_\odot \cdot {\rm pc}^{-3}}, \label{eq:P_surv}
\end{align}
where the $\Phi$-dependence appears through the MC density $\rho_\mathrm{MC}(\Phi)$ in Eq.~\eqref{eq:rho_mc}.
Using Eq.~\eqref{eq:rho_mc}, we can find the survival probability in terms of $\Phi$.
$\mathcal{P}_\mathrm{surv}(\Phi)$ is shown in Fig.~\ref{fig:P_surv}.

\bibliographystyle{JHEP}
\bibliography{accAS}

\end{document}